\documentclass[%
 longbibliography,
 prapplied,%
 amsmath,amssymb,
 reprint,%
 superscriptaddress
,floatfix
]{revtex4-2}

\usepackage{graphicx}
\usepackage{siunitx}
\usepackage[utf8]{inputenc}
\usepackage{amsmath}
\usepackage{amsfonts}
\usepackage{amssymb}
\usepackage{natbib}
\usepackage{xcolor}
\usepackage{booktabs}
\usepackage{multirow}
\usepackage{array}
\usepackage{braket}
\usepackage{soul}
\usepackage{floatrow}
\graphicspath{{Figs/}}
\usepackage{hyperref}
\hypersetup{
      filecolor=blue,
      urlcolor=blue
      colorlinks=true,
      linkcolor=blue,
      citecolor=blue} 

\makeatletter
\newcommand{\manualsublabel}[3]{(#2)\def\@currentlabel{\ref{#3}(#2)}\label{#1}}

\newfloatcommand{capbtabbox}{table}[][\FBwidth]
\DeclareUnicodeCharacter{FB01}{}
\makeatother

\begin{document}

\title{Site-resolved magnon and triplon dynamics on a programmable quantum dot spin ladder}
\date{\today}

\author{Pablo Cova Fari\~na}
\thanks{These authors contributed equally to this work}
\email{p.covafarina@tudelft.nl}
\affiliation{QuTech and Kavli Institute of Nanoscience, Delft University of Technology, 2600 GA Delft, The Netherlands}
\author{Daniel Jirovec$^*$}
\affiliation{QuTech and Kavli Institute of Nanoscience, Delft University of Technology, 2600 GA Delft, The Netherlands}
\thanks{These authors contributed equally to this work}
\author{Xin Zhang$^*$}
\affiliation{QuTech and Kavli Institute of Nanoscience, Delft University of Technology, 2600 GA Delft, The Netherlands}
\author{Elizaveta Morozova$^*$}
\affiliation{QuTech and Kavli Institute of Nanoscience, Delft University of Technology, 2600 GA Delft, The Netherlands}
\author{Stefan D. Oosterhout}
\affiliation{Netherlands Organisation for Applied Scientific Research (TNO), 2628 CK Delft, The Netherlands}
\author{Stefano Reale}
\affiliation{QuTech and Kavli Institute of Nanoscience, Delft University of Technology, 2600 GA Delft, The Netherlands}
\author{Tzu-Kan Hsiao}
\affiliation{Department of Physics, National Tsing Hua
University, Hsinchu 30013, Taiwan}
\author{Giordano Scappucci}
\affiliation{QuTech and Kavli Institute of Nanoscience, Delft University of Technology, 2600 GA Delft, The Netherlands}
\author{Menno Veldhorst}
\affiliation{QuTech and Kavli Institute of Nanoscience, Delft University of Technology, 2600 GA Delft, The Netherlands}
\author{Lieven M. K. Vandersypen}
\email{L.M.K.Vandersypen@tudelft.nl}
\affiliation{QuTech and Kavli Institute of Nanoscience, Delft University of Technology, 2600 GA Delft, The Netherlands}

\date{\today}

\begin{abstract}
Quasi-particle dynamics in interacting systems in the presence of disorder challenges the notion of internal thermalization, but proves difficult to investigate theoretically for large particle numbers. Engineered quantum systems may offer a viable alternative, as witnessed in experimental demonstrations in a variety of physical platforms, each with its own capabilities and limitations. Semiconductor gate-defined quantum dot arrays are of particular interest since they offer both a direct mapping of their Hamiltonian to Fermi-Hubbard and Heisenberg models and the in-situ tunability of (magnetic) interactions and onsite potentials. In this work, we use an array of germanium quantum dots to simulate the dynamics of both single-spin excitations (magnons) and two-spin excitations (triplons). We develop a methodology that combines digital spin qubit operations for state preparation and readout with analog evolution under the full system Hamiltonian. Using these techniques, we can reconstruct quantum walk plots for both magnons and triplons, and for various configurations of Heisenberg exchange couplings. We furthermore explore the effect of single-site disorder and its impact on the propagation of spin excitations. The obtained results can provide a basis for simulating disorder-based solid-state phenomena such as many-body localization.

\end{abstract}

\maketitle

\section{Introduction}

The observation of correlated many-body phenomena is one of the most promising applications of quantum simulators \cite{feynman_simulating_1982,lloyd_universal_1996,georgescu_quantum_2014}. For example, studying the dynamics of spin excitations under a given many-body spin model can provide insight into the system's underlying Hamiltonian and its magnetic properties. A fascinating open question pertains to the properties of isolated quantum systems in the presence of disorder, which competes with interactions and can significantly alter the propagation of spin excitations or the thermalization properties of the system \cite{abanin_colloquium_2019}. While different simulator platforms have implemented this Hamiltonian class \cite{schreiber_observation_2015,kondov_disorder-induced_2015,smith_many-body_2016,choi_exploring_2016,roushan_spectroscopic_2017,xu_emulating_2018,morong_observation_2021,mehta_down-conversion_2023}, a rigorous study of such models ideally requires precise tunability of both spin-spin interactions and single-site disorder, as well as reliable initialization and readout of the spin states with single-site resolution. When comparing different quantum simulation platforms, atomic systems \cite{bloch_quantum_2012,blatt_quantum_2012,kaufman_quantum_2021,cornish_quantum_2024} can reach large sizes but are often limited in their local control capabilities, while solid-state-based systems \cite{kjaergaard_superconducting_2020,choi_colloquium_2019,hensgens_quantum_2017} offer a high degree of tunability but need to tackle other challenges such as crosstalk, homogeneity and scalability.

In this context, gate-defined semiconductor quantum dots provide a versatile platform to perform quantum simulations of Fermi-Hubbard \cite{barthelemy_quantum_2013,hensgens_quantum_2017,dehollain_nagaoka_2020,kiczynski_engineering_2022,hsiao_exciton_2024} and Heisenberg-type \cite{van_diepen_quantum_2021,wang_probing_2023} physics. Precise control of onsite energies and local couplings is routinely achieved through voltages applied on dedicated electrodes. Additionally, quantum dot arrays naturally emulate the Heisenberg model at half filling, making them a promising candidate for simulating spin physics. Among the available semiconductor platforms, Ge/SiGe quantum dots have been rapidly scaled up in recent years \cite{van_riggelen_two-dimensional_2021,jirovec_singlet-triplet_2021,scappucci_germanium_2021,borsoi_shared_2024,wang_operating_2024,john_two-dimensional_2024}, enabling the fabrication of 2D quantum dot arrays and device sizes approaching the many-body regime. Moreover, significant variations of the $g$-tensors of hole spins in germanium quantum dots and their principal axes are commonly observed, which has enabled qubit addressability for spin-qubit experiments~\cite{hendrickx_four-qubit_2021,john_two-dimensional_2024}. In the context of this work, this $g$-tensor variability also provides a native realization of random single-site disorder. Finally, individual spin exchange couplings, arising due to wavefunction overlap between neighboring dots, can be quickly turned on or off by gate voltage pulses, in principle allowing to observe the evolution of the system after a Hamiltonian quench.

\begin{figure*}
    \centering
    \includegraphics[width=0.95\textwidth]{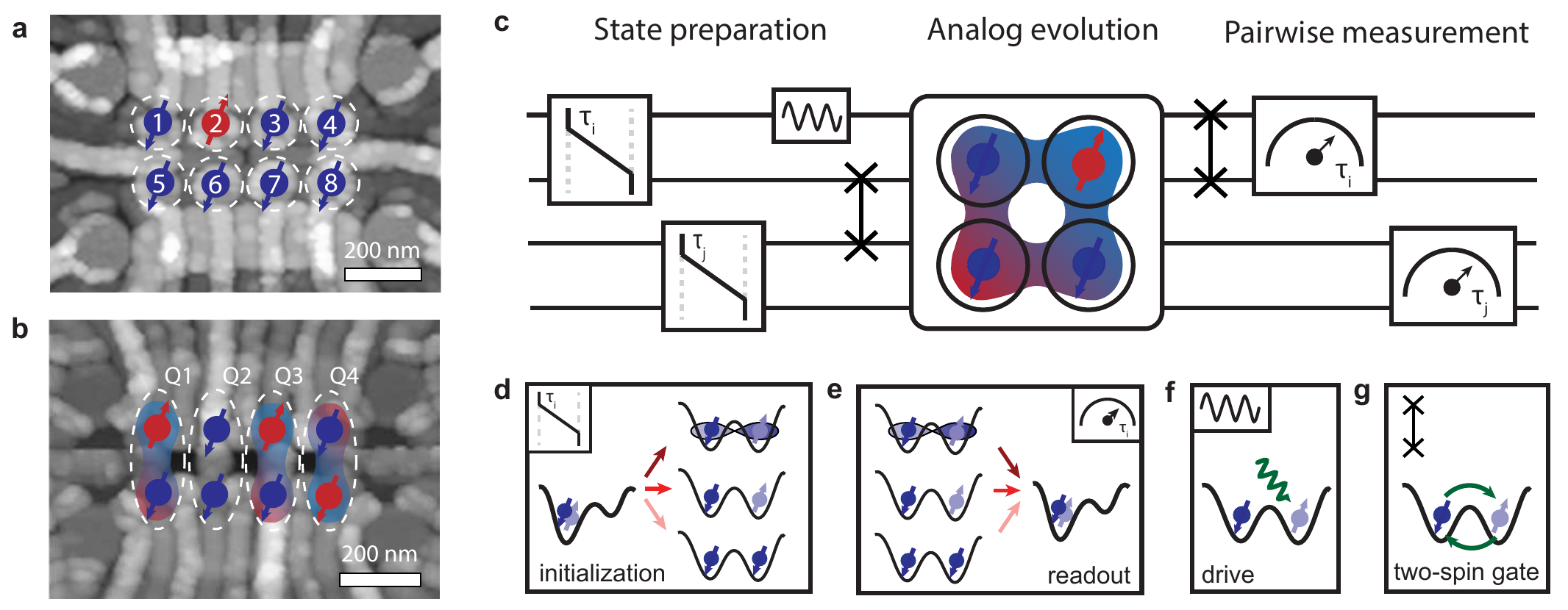}
    \caption{Atomic-force microscopy images of the two germanium 2$\times$4 quantum dot devices used in this work. Spins are schematically depicted to highlight the two different types of encoding: (a) single-spin and (b) singlet-triplet. (c) Schematics of a digital-analog circuit: During the state preparation and readout phases, performed in the computational (Zeeman) basis, a combination of ramp times $\tau_i$ of the interdot detunings with single- and two-qubit gates allows the preparation of a variety of states as well as the extraction of single-site spin information. During the analog evolution part, spin-exchange couplings can be diabatically or adiabatically turned on, resulting in spin dynamics under the system's native Hamiltonian, as described in the main text. Schematic depiction of (d) initialization (see main text), (e) readout (see main text), (f) single-spin drive using microwaves and (g) two-spin gates based on Heisenberg exchange utilized in this work.}
    \label{fig:intro}
\end{figure*}

In this work, we explore the dynamics of two kinds of spin excitations, namely magnons and triplons, and their propagation for different exchange coupling configurations in a semiconductor 2$\times$4 quantum dot ladder. For this purpose, we use an analog-digital experimental framework \cite{parra-rodriguez_digital-analog_2020,andersen_thermalization_2025}. For initialization and readout,  we combine quantum gates and transitions with a controlled degree of adiabaticity, which allows us to prepare spin excitations on arbitrary sites and track their position over time. In between, we allow the system to evolve under its native Hamiltonian. We first investigate the dynamics of single-spin excitations, or magnons, under different coupling topologies and disorder-to-interaction ratios, for which we leverage on a recently developed crosstalk mitigation protocol for exchange couplings in dense quantum dot arrays \cite{jirovec_exchange_2025}. We develop methods to prepare a magnon in any site of the ladder and probe the presence of a magnon in a site-resolved manner, allowing us to reconstruct quantum walks~\cite{fukuhara_microscopic_2013,jurcevic_quasiparticle_2014,yan_strongly_2019}. Next, we study the propagation of two-spin excitations, or triplons~\cite{dagotto_superconductivity_1992,giamarchi_quantum_2010,nawa_triplon_2019}, along a chain of dimerized sites, building on previous work on singlet-triplet qubit control across a quantum dot ladder~\cite{zhang_universal_2024}. We demonstrate a method to initialize a triplon in any site, fully and independently tune interactions and on-site disorder, and detect the presence or absence of a triplon on three sites in a single-shot manner. 

\section{Device and mode of operation}

Fig. \ref{fig:intro}a and b show two atomic-force microscopy images of the two Ge/SiGe 2$\times$4 quantum dot devices used in this work. While Fig. \ref{fig:intro}a corresponds to the device used for the magnon propagation experiments (Sections \ref{sec:initread} and \ref{sec:qwalks}), we use the device in Fig. \ref{fig:intro}b to explore triplon propagation in the singlet-triplet basis (Section \ref{sec:S-T$^{-}$_triplons}). On this same device, universal control of four singlet-triplet qubits  was previously demonstrated \cite{zhang_universal_2024}. 

The device of Fig. \ref{fig:intro}a incorporates a few small but relevant changes with respect to the previous version \cite{zhang_universal_2024}. First, the barrier gates are patterned in the first metal layer to increase their lever arm and, consequently, the ability to tune the exchange interaction strengths. Additionally, the charge sensors, used for tuning and readout purposes, are not tunnel coupled to the device while maintaining a high capacitive coupling, which improved the sensing dot tunability (see Fig. \ref{fig:device_layout} for details on the device layout).

We operate both devices in a mixed analog-digital fashion (Figs. \ref{fig:intro} c-g). We rely on pairwise spin initialization and readout based on Pauli-spin blockade (PSB) by controlling the ramp speed between the (2,0) and (1,1) states (see Section ~\ref{sec:initread}), where ($n_l,n_r$) represent the number of charges in the left and right dot of a quantum dot pair. Although the key element of these experiments is the analog evolution part, we also employ single- and two-qubit gates as a means to quickly and reliably prepare any desired initial state relevant for this work, and to access additional readout bases. Note that for this particular work, microwave driving is only used for qubit spectroscopy, since we operate at a low magnetic field of \SI{10}{mT}, where microwave-driven gates are slow~\cite{bulaev_electric_2007,hendrickx_sweet-spot_2024,froning_ultrafast_2021,mauro_geometry_2024}.

\section{Spin Hamiltonians} 

\renewcommand{\arraystretch}{1.1}

The Hamiltonian of an array of quantum dots with a single spin per site and an external magnetic field $B$ is given by a Heisenberg model with exchange interactions $J_{ij}$ and site-dependent $g$-factors $g_i$:

\begin{equation}
    H_{0} = \sum_{\langle i,j \rangle} J_{ij} \mathbf{S_i} \cdot \mathbf{S_j} + \mu_B B \sum_i g_i S_i^z,
    \label{eq:spinham}
\end{equation}

where the first sum runs only over nearest neighbors $i$ and $j$, $\mathbf{S_i}$ is the vector of spin operators $(S_i^x, S_i^y, S_i^z) = \frac{\hbar}{2} (\sigma_i^x, \sigma_i^y, \sigma_i^z)$, using the Pauli matrices for each spin $i$. For holes in germanium quantum dots, this Hamiltonian is only an approximation, neglecting the tensorial nature of $J_{ij}$ and $g_i$ \cite{hetenyi_exchange_2020}. These terms, and additional intrinsic spin-orbit spin flip components, are typically captured by an additional term of the form $\sum_i \Delta_{ST,i} S_{x,i}$, with $\Delta_{ST}$ the spin-orbit strength, a term we will exploit for initialization and readout but ignore during the analog evolution. Despite these simplifications, the Hamiltonian of Eq.~\ref{eq:spinham} suffices to model the experimental data revealing magnon propagation, see also the discussion in Section \ref{sec:disc}. Important for studies of spin dynamics, the spin exchange interactions are individually tunable through the gate voltage that controls the interdot tunnel coupling. Also the individual $g$-tensors have been demonstrated to be electrically tunable by modulating the shape of the quantum dot wavefunction~\cite{scappucci_germanium_2021,jirovec_singlet-triplet_2021,mauro_geometry_2024,wang_modeling_2024}. However, this tunability is often hard to predict in practice since it is highly dependent on microscopic device details.

Next we consider an array of singlet-triplet qubits, where each rung $i$ of the ladder encodes a single qubit in the singlet-triplet basis (see Fig. \ref{fig:intro}b). When projected onto the S-T$^{-}$ subspace, the coupling Hamiltonian of two singlet-triplet qubits can be mapped to \cite{zhang_universal_2024}:

\begin{equation} \label{eq:stham}
\begin{split}
    H_{ST} &= \frac{1}{4}\sum_{\langle i,j \rangle} J_{ij}^\parallel \left[\sigma_i^x \sigma_j^x + \sigma_i^y \sigma_j^y
    + \frac{1}{2} \left(\sigma_i^z - I\right) \left(\sigma_j^z - I\right) \right]\\ &+ \frac{1}{2}\sum_i (\bar{E}_{z,i}-J_i^\perp) \sigma_i^z,
\end{split}
\end{equation}

where $J_{ij}^\parallel$ and $J_i^\perp$ are the exchange couplings between different rungs $i,j$ and within each rung $i$, respectively.
In this case, single-site energy terms become programmable by means of $J_i^\perp$. Note that in Eq.~\ref{eq:stham} we also neglect the spin-orbit coupling (SOC) terms \cite{zhang_universal_2024}, which would result in deviations from the isotropic behavior. This is only valid far from the spin-orbit induced avoided crossings, which is the case during the analog evolution part. In contrast, for initialization and readout, we explicitly pulse to the spin-orbit induced S-T$^{-}$ avoided crossing to achieve single-qubit rotations (see Section \ref{sec:S-T$^{-}$_triplons}). 

Interestingly, the interaction terms of Eqs. \ref{eq:spinham} and \ref{eq:stham} realize two different spin XXZ Hamiltonians of the form 
\begin{equation}
    H_{\text{int}} \propto \sum_{\langle i,j \rangle} \left[ \sigma_i^x\sigma_j^x +  \sigma_i^y\sigma_j^y + \Delta \sigma_i^z\sigma_j^z \right],
\end{equation}
with $\Delta$ the anisotropy parameter ~\cite{fukuhara_microscopic_2013}. While Eq. \ref{eq:spinham} is a Heisenberg or XXX Hamiltonian  with $\Delta = 1$, in the triplon case we find $\Delta = 0.5$ (Eq. \ref{eq:stham}). Table \ref{tab:comparison} summarizes both encodings.

\begin{table}[t]
\centering
\begin{tabular}{>{\centering\arraybackslash}m{2.5cm} >{\centering\arraybackslash}m{3cm} >{\centering\arraybackslash}m{2.5cm}}
\toprule
\textbf{Encoding}                               & \textbf{Single-spin}                                                                                    & \textbf{Singlet-triplet} \\ \midrule
Number of sites for $N$ dots      & $N$                                                                                                                & $N/2$                               \\ \midrule
Interaction Hamiltonian (neglecting SOC)                 & XXX ($\Delta = 1$)                                                                                                              & XXZ ($\Delta = 0.5$)                           \\ \midrule
Interaction tunability                  & Yes                                                                                                              & Yes                               \\ \midrule
Disorder source                         & $g$-factor variability                                                                                                         & $\bar{E}_{z}-J^\perp$                         \\ \midrule
Disorder tunability                     & \begin{tabular}[c]{@{}c@{}}Global (B-field)\\ Limited local tunability\\ (g-tensor modulation)\end{tabular} & Yes                               \\ \midrule
Site-resolved readout                   & PSB + SWAP                                                                                                       & PSB                               \\ \bottomrule
\end{tabular}
\caption{Comparison between single-spin and singlet-triplet encodings.}
\label{tab:comparison}
\end{table}

\section{State initialization and readout} \label{sec:initread}

\begin{figure*}
    \centering
    \includegraphics[width=\textwidth]{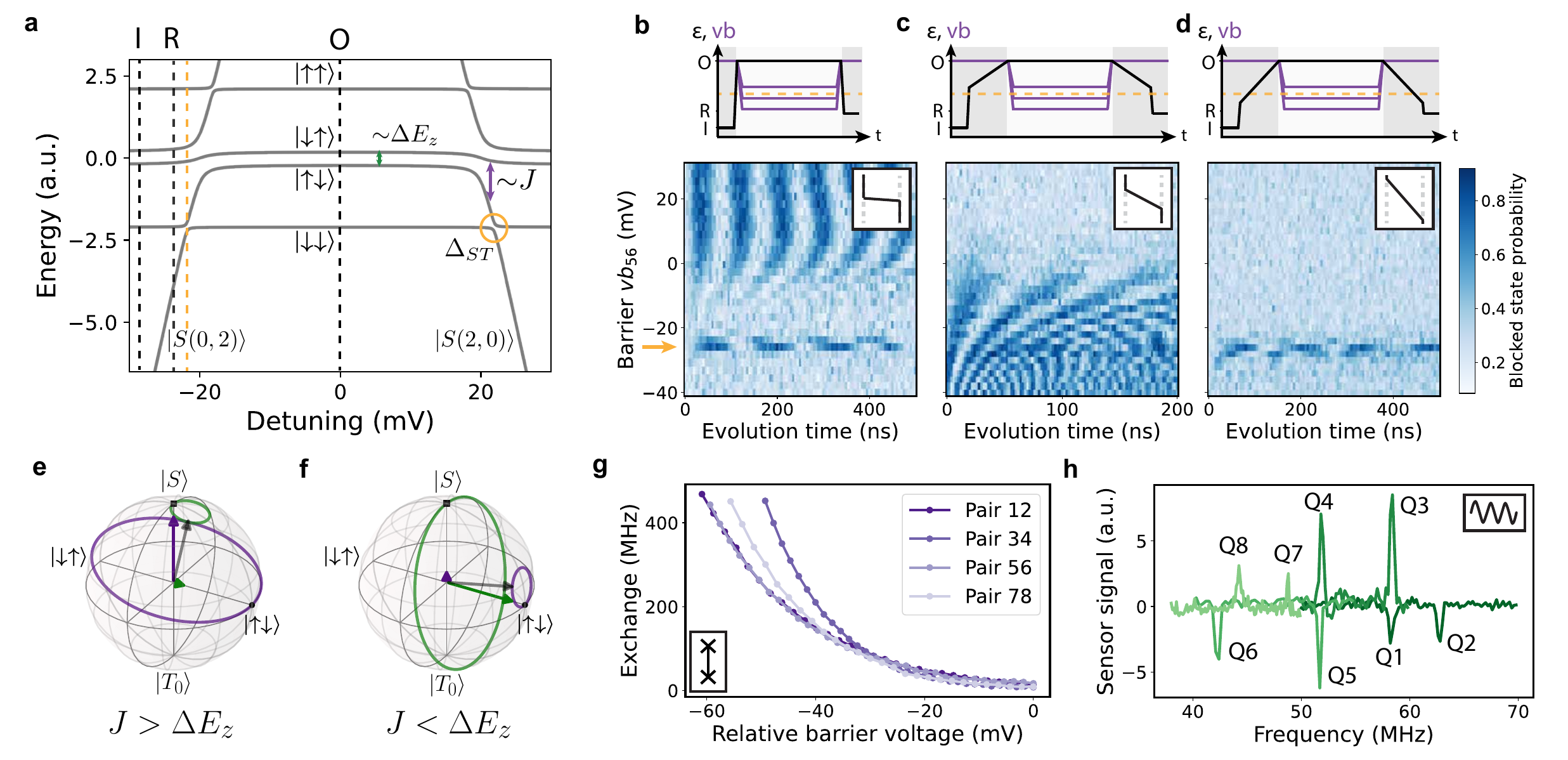}
    \caption{Pairwise initialization, readout, characterization and control of hole spins. (a) Energy diagram for a pair of spins. The black dashed lines represent the initialization (I), operation (O) and readout (R) points, respectively. The S-T$^{-}$ anticrossing is highlighted in yellow. (b) Experimentally observed dynamics of an initial state $\ket{S}$ as a function of virtual barrier voltage vb$_{56}$, which controls the exchange interaction strength, for quantum dots 5-6. A singlet is initialized by pulsing with a fast \SI{4}{ns} ramp from the initialization to the operation point, and read out by pulsing fast from the operation to the readout point. S-T$^0$ oscillations are visible at low exchange (positive barrier voltage). Additional oscillations at high exchange (negative barrier voltage) are consistent with an evolution at the S-T$^{-}$ anticrossing. (c) Experimentally observed dynamics of an initial state $\ket{\uparrow\downarrow}$ as a function of barrier voltage for quantum dots 5-6. The state $\ket{\uparrow\downarrow}$ is initialized by pulsing with a fast \SI{4}{ns} ramp from the initialization point over the S-T$^{-}$ anticrossing, and then slowly ramping ($\tau_{\text{ramp}} \sim$ \SI{1}{\micro s}) to the operation point. For readout, this pulse scheme is inverted. Exchange oscillations are visible at high values of $J$. (d) Experimentally observed dynamics of an initial state  $\ket{\downarrow\downarrow}$ as a function of barrier voltage for quantum dots 5-6. The state $\ket{\downarrow\downarrow}$ is initialized by pulsing with a fast \SI{4}{ns} ramp from the initialization point until before the S-T$^{-}$ anticrossing, and then slowly ramping ($\tau_{\text{ramp}} \sim$ \SI{1}{\micro s}) to the operation point. For readout, this pulse scheme is inverted. The only observed dynamics is at the S-T$^{-}$ anticrossing. The diagrams above the data schematically show the corresponding pulse schemes for both detuning (black lines) and barrier voltage (purple lines). The gray shaded regions correspond to initialization and readout. (e) Bloch sphere depiction of the time evolution of a singlet and a $\ket{\uparrow\downarrow}$ state for an interaction-dominated Hamiltonian ($J > \Delta E_z$). (f) Bloch sphere depiction of the time evolution of a singlet and a $\ket{\uparrow\downarrow}$ state for a disorder-dominated Hamiltonian ($J < \Delta E_z$). (g) Exchange coupling tunability as a function of barrier gate voltage, extracted from Fig. \ref{fig:magnon_all}. We obtain interaction strengths of up to 500 MHz and down to well below 1 MHz for each dot pair, highlighting the excellent individual tunability of the exchange couplings. In particular, we can comfortably reach the $J \gg \Delta E_z$ and $J \ll \Delta E_z$ regimes. (h) Zeeman energy extraction through EDSR. For each pair of spins, the resonances are assigned depending on the driving gate (we flipped half of the signals for clarity). This sets the single-particle disorder landscape. We find that neighboring spins differ in frequency by 5-\SI{15}{MHz} at a magnetic field of \SI{10}{mT}.}
    \label{fig:initread}
\end{figure*}

Fig. \ref{fig:initread}a shows the typical energy diagram of a germanium double quantum dot \cite{jirovec_singlet-triplet_2021,jirovec_dynamics_2022} as a function of detuning $\varepsilon_{ij} = \frac{1}{2}(\mu_j - \mu_i)$, with $\mu_i$ the electrochemical potentials of two neighboring dots $i$ and $j$. This energy diagram is simulated using the lowest energy levels of a two-site Fermi-Hubbard model (see Eq. \ref{eq:FH_dd} of the Supplementary Material). The detuning axis is chosen such that the total charge occupation of the system is conserved.  At the symmetry point $\varepsilon_{ij} = 0$, there is one charge per quantum dot (charge state (1,1)) and, at small tunnel coupling values $t_{ij}$, the spin eigenstates are the four product states $\{\ket{\downarrow\downarrow}, \ket{\uparrow\downarrow},  \ket{\downarrow\uparrow},\ket{\uparrow\uparrow}\}$. Away from $\varepsilon = 0$ (leaving out the subscripts for simplicity), the wavefunction overlap between the two charges increases, resulting in an increased exchange coupling $J$. The spin-orbit induced anticrossing between the singlet and triplet branches occurs at $J = \bar{E_z}$. Finally, at very positive or negative detuning, with absolute values larger than the quantum dot charging energies, both charges sit on the same quantum dot, and the ground state of the system is a singlet (S(2,0) or S(0,2), respectively).

The initialization procedure is derived from this energy diagram. Starting from S(2,0), we ramp to the symmetry point at $\varepsilon = 0$ with a ramp time $\tau$. Crucially, the state we initialize depends on the ramp rate $v = \frac{dE}{d\tau}$ and two relevant energy scales: the S-T$^{-}$ anticrossing size $\Delta_{ST}$ and the Zeeman energy difference between the two dots $\Delta E_z$ \cite{harvey-collard_spin-orbit_2019,kelly_identifying_2025}. Using realistic experimental parameters, we simulate the initialized states as a function of ramp speed (Fig. \ref{fig:init_sim}) and recognize three distinct regimes. At diabatic ramp speeds $\hbar v \gg \Delta E_z^2$, $\hbar v \gg \Delta_{ST}^2$, the initial spin state is conserved, resulting in the state $S(1,1)$. In contrast, at adiabatic speeds, the ground state of the system $\ket{\downarrow\downarrow}$ is initialized. At intermediate speeds, when the ramp speed is adiabatic only with respect to the Zeeman energy difference, the state $\ket{\uparrow \downarrow}$ is initialized, with the spin-up corresponding to the dot with the lowest $g$-factor. 

The same reasoning applies to readout, which is based on Pauli spin blockade \cite{ono_current_2002,engel_measurement_2004,petta_coherent_2005,seedhouse_pauli_2021,lundberg_non-symmetric_2024,nurizzo_complete_2023}. The ramp rate from the symmetry point to the readout point determines which two-spin state is unblocked and therefore distinguished from the three other, orthogonal, spin states. While a diabatic ramp will result in an unblocked S(1,1) state, the state $\ket{\downarrow\downarrow}$ ($\ket{\uparrow\downarrow}$) is unblocked for adiabatic (intermediate) ramp speeds. To further improve the initialization and readout fidelities, we make use of composite ramps as depicted in Figs. \ref{fig:initread}b-d and detailed in Fig. \ref{fig:csd_init}. Note that more complex ramps could be implemented to further optimize initialization and readout \cite{meinersen_quantum_2024}.

We experimentally demonstrate the initialization, readout and exchange control capabilities in Fig.~\ref{fig:initread}b-d. Each of the panels shows the evolution of a two-spin initial state for quantum dot pair 56 as a function of its corresponding virtual barrier \cite{jirovec_exchange_2025} voltage vb$_{56}$ for diabatic, intermediate and adiabatic initialization and readout ramps, respectively. The dynamics in Figs.~\ref{fig:initread}b and c can be understood through a Bloch sphere representation in the S-T$^0$ basis~\cite{jirovec_singlet-triplet_2021} (Figs.~\ref{fig:initread}e and f). At positive barrier voltages, the exchange interaction between dots is minimized ($J \ll \Delta E_z$, disorder-dominated regime) and the eigenstates are well approximated by the eigenstates of the Zeeman Hamiltonian, i.e. the product states $\{\ket{\downarrow \downarrow},\ket{\uparrow \downarrow},\ket{\downarrow \uparrow},\ket{\uparrow \uparrow}\}$. An initial $\ket{S}$ state, being an equal superposition of $\ket{\uparrow \downarrow}$ and $\ket{\downarrow \uparrow},$ should therefore oscillate with maximum visibility, while an initial $\ket{\uparrow \downarrow}$ state should not oscillate. The opposite behavior is expected at negative barrier voltages, i.e. high exchange couplings ($J \gg \Delta E_z$, interaction-dominated regime). In either of the two cases, the $\ket{\downarrow \downarrow}$ state should not evolve since it is an eigenstate of both the Zeeman and the Heisenberg Hamiltonians. Figs. \ref{fig:initread}b-d show the expected evolution patterns, highlighting not only the initialization and readout capability in different bases, but also the excellent control over the ratio between interactions and Zeeman energy differences. Note that at interaction strengths $J = \bar{E_z}$, we observe an extra set of oscillations for all three initialized states, consistent with the position of the S-T$^{-}$ anticrossing, and validated through simulations (see Fig. \ref{fig:osc_all_sim}).

We demonstrate this behavior for all four spin pairs individually (see Fig. \ref{fig:st_all} and \ref{fig:magnon_all}) and extract exchange interaction values of up to \SI{500}{MHz} and down to well below 1 MHz within a comfortable dynamical range in the barrier gate voltages of $< \SI{100}{mV}$ (Fig. \ref{fig:initread}g). Note that the onset of oscillations for the $\ket{\uparrow \downarrow}$ plots depends on the $\Delta E_z$ value of each pair, which differs due to the difference in $g$-factors. This well-known behavior for two spins forecasts a more generic behavior, where a spin excitation which can propagate along an array of spins under the influence of exchange interactions $J$ becomes localized when disorder in the local Zeeman energies $\Delta E_z$ exceeds spin exchange. 

We make use of microwave driving to extract the individual Zeeman energies via electric-dipole spin resonance (EDSR) (Fig. \ref{fig:initread}h). We obtain the resonance frequencies by initializing a product state for each pair and driving with different plunger gates to pinpoint which resonance line corresponds to which spin, or by measuring the exchange splitting of the resonance lines as a function of neighboring barrier voltages (see Fig. \ref{fig:g_factor_mod}). For the applied in-plane field of \SI{10}{mT}, the Zeeman energy differences are of the order of 5-\SI{15}{MHz}, while the Zeeman energies correspond to $g$-factors of $0.30 < g_i < 0.44$, slightly higher than the values reported in the literature \cite{hendrickx_four-qubit_2021}. This could be caused by misalignment of the external magnetic field, leading to a small out-of-plane field component.

\section{SWAP gates, single-site readout and quantum walks} \label{sec:qwalks}

\begin{figure*}
    \centering
    \includegraphics[width=\textwidth]{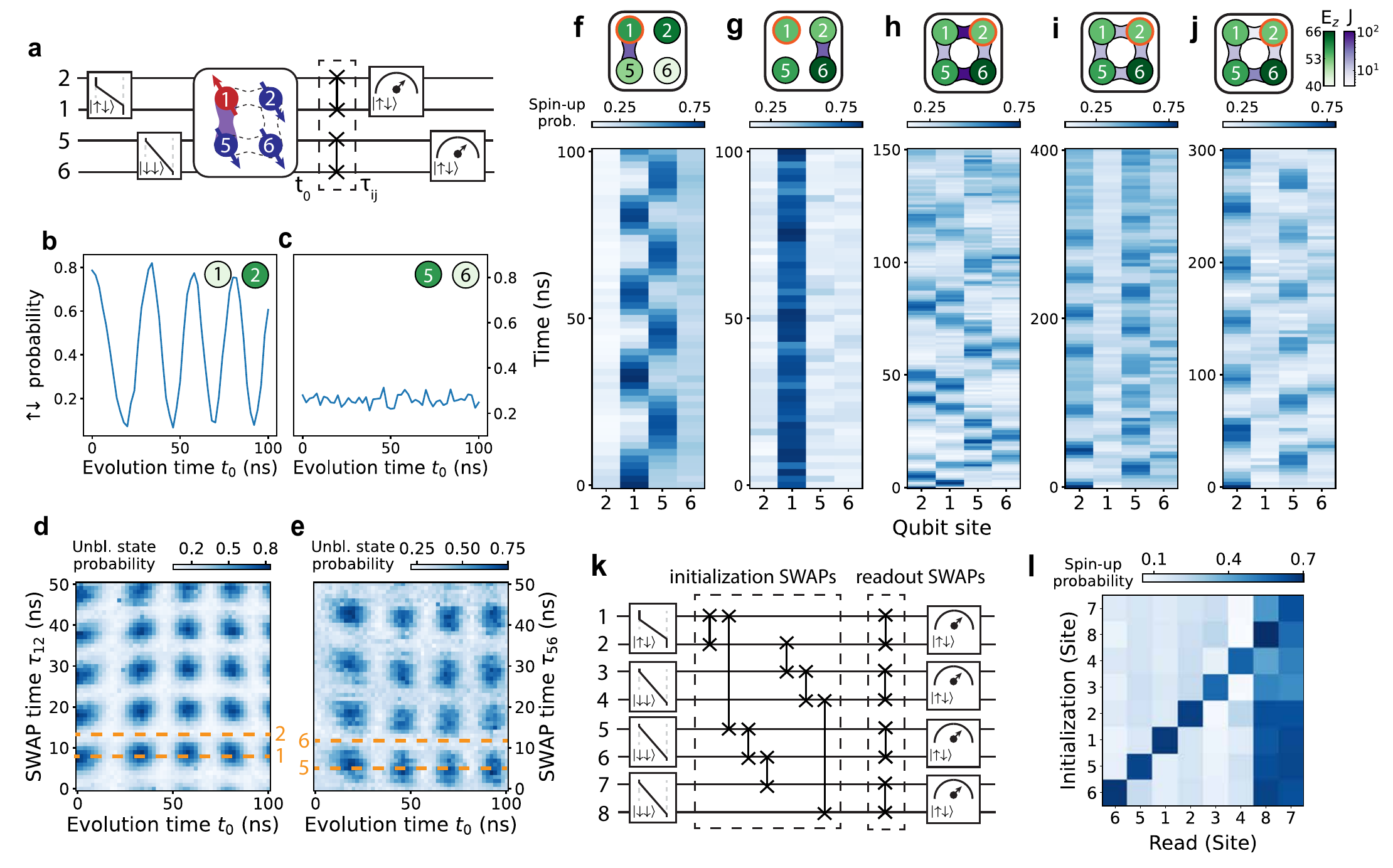}
    \caption{Magnon propagation for the left half of the 2$\times$4 array with various coupling configurations. (a) Circuit diagram for simultaneous calibration of SWAP times $\tau_{12}$ and $\tau_{56}$. For this particular experiment, the only exchange that is switched on during the analog evolution is $J_{15}$. $\ket{\uparrow\downarrow}$-readout allows to obtain the spin-up probability of the quantum dot with the lowest $g$-factor for each pair. (b) $\ket{\uparrow\downarrow}$ probability for quantum dot pair 1-2 as a function of evolution time $t_0$, recorded after running the quantum circuit in panel (a) without the SWAP operations. (c) $\ket{\uparrow\downarrow}$ probability for quantum dot pair 5-6 as a function of evolution time $t_0$, recorded after running the quantum circuit in panel (a) without the SWAP operations. (d) and (e) Unblocked state probability for (d) pair 1-2 and (e) pair 5-6 after executing the quantum circuit of panel (a), as a function of $t_0$ and the SWAP times as indicated. From these 2D maps, we extract the SWAP times which correspond to spin-up readout for each of the four spins (indicated by the dashed orange lines). (f) Four-spin quantum walk corresponding to the circuit in panel (a) and extracted from plots (d) and (e), showing evolution of the spin-up probability for the sites 2, 1, 5 and 6 as a function of $t_0$. For this and the following quantum walk plots, the border of the quantum dot with the initial spin-up is highlighted in orange, the magnitude of the Zeeman energies is color-coded in green, and the exchange interaction strengths in purple. (g) Four-spin quantum walk in a 2$\times$2 array with $J_{26} \approx 40$ MHz and the other exchanges switched off. (h) Four-spin quantum walk in a configuration with $J_{12} \approx J_{56} \approx 100$ MHz  and  $J_{15} \approx J_{26} \approx 25$ MHz. (i) Four-spin quantum walk in a 2$\times$2 ring with homogeneous exchange couplings ($J_{ij} \approx$ \SI{21}{MHz}). The excitation mainly oscillates between dots 2 and 5, a beating pattern is additionally observed. (j) Four-spin quantum walk in a 2$\times$2 ring with disorder in the $J_{ij}$ values, with barrier voltages slightly tuned away from the homogeneous exchange point. A rather regular oscillation pattern mainly between dots 2 and 5 is observed. (k) Circuit diagram depicting the preparation of any initial state $\ket{i}$, with a single spin-up initialized in site $i$. After preparing the state $\ket{2}$ adjusting the speed of the initialization ramps, the single excitation is swapped around the array. Finally, the resulting state is read out. Initialization and readout happen sequentially and not simultaneously (see depiction of panel (a), but are drawn here on top of each other for compactness. (l) Spin-up probability for the full 8-site array as a function of readout site and target initial state.}
    \label{fig:swaps}
\end{figure*}

Building on the tools for initialization and readout showcased in section~\ref{sec:initread}, we can study the propagation of single-spin excitations or magnons in extended quantum dot arrays. The three different readout types outlined above allow us to discriminate a single unblocked two-spin state ($\ket{S}$, $\ket{\uparrow\downarrow}$ or $\ket{\downarrow\downarrow}$ respectively) from the states in its orthogonal subspace. This does not directly translate to single-spin probabilities. For instance, reading out at an intermediate ramp speed yields the probability of measuring $\ket{\uparrow\downarrow}$, but the readout signal would not distinguish between the states $\ket{\downarrow\uparrow}$, $\ket{\downarrow\downarrow}$, and $\ket{\uparrow\uparrow}$ which are all blocked. However, apart from the $\ket{\uparrow\downarrow}$ and $\ket{\downarrow\downarrow}$ readouts demonstrated in section \ref{sec:initread}, we can also read out the orthogonal state $\ket{\downarrow\uparrow}$ by implementing a SWAP gate before ramping to the readout point (Fig. \ref{fig:swaps}a). Since the exchange control permits to reach $J \gg \Delta E_z$ for all dot pairs, the SWAP gate is a native two-qubit gate of the system, achievable by pulsing the barrier gates for a time $\tau_{\text{SWAP}} = \pi/ J$. In this way, readout of three orthogonal two-spin states is possible from which the fourth can be inferred through normalization, allowing us to reconstruct all single-spin probabilities. Importantly, if only a single excitation is present in the array, only two of the three readouts are necessary since $P(\ket{\uparrow\uparrow}) = 0$. In this case, $P(\ket{\uparrow_i}) = P(\ket{\uparrow_i\downarrow_j})$ and $P(\ket{\uparrow_j}) = P(\ket{\downarrow_i\uparrow_j})$. This is the case for all quantum walk data shown in Fig. \ref{fig:swaps}.

In Fig. \ref{fig:swaps}a-f, we show a method for simultaneously calibrating two SWAP gates in a four-spin system and to obtain site-resolved quantum walk plots. For this example, we initialize the spin state $\ket{1} = \ket{\uparrow_1\downarrow_2\downarrow_5\downarrow_6}$. During the analog evolution, the exchange coupling $J_{15}$ is activated for a period of time $t_0$, which should yield a simple oscillatory pattern between spins 1 and 5. After the evolution, we either perform $\ket{\uparrow\downarrow}$-readout directly on pairs 1-2 and 5-6, or we activate the exchanges $J_{12}$ and $J_{56}$ for a variable time $\tau_{12}$ and $\tau_{56}$ before readout. This is depicted as a quantum circuit in Fig. \ref{fig:swaps}a. For this four-spin experiment, the $g$-factor landscape corresponds to Fig. \ref{fig:initread}h. This means that the dots with the lowest $g$-factors are dot 1 for the top pair and dot 6 for the bottom pair.

When no SWAP gate is applied before readout (Figs. \ref{fig:swaps}b and c), we observe an oscillating pattern for the top pair and a constant low probability for the bottom pair, consistent with the readout of spins 1 and 6. Figs. \ref{fig:swaps}d and e show the measured spin probabilities as a function of time $t_0$ and the SWAP time before readout. At SWAP times $\tau_{ij} = 2\pi/J_{\text{SWAP},ij}$, the line cuts correspond to Figs.~\ref{fig:swaps}b and c. In contrast, when $\tau_{ij} = \pi/J_{\text{SWAP},ij}$, we obtain the readout of dots 2 and 5, respectively, resulting in oscillations of the bottom pair and a low constant signal for the top pair. From this 2D map, we extract the correct SWAP times and reconstruct the four-spin quantum walk in Fig. \ref{fig:swaps}f, showing spin oscillations between sites 1 and 5, as expected. Notably, the 2D maps in Figs.~\ref{fig:swaps}d and e also contain readout probabilities of other two-spin operators for intermediate SWAP times, which might prove useful to directly obtain other observables of interest~\cite{impertro_local_2024}.

This methodology allows us to experimentally observe quantum walks with different coupling topologies, where the interplay between disorder, interactions and evolution time becomes crucial. As a first example and in contrast to the free spin propagation of Fig. ~\ref{fig:swaps}f,  Fig.~\ref{fig:swaps}g shows an initial magnon remaining localized in quantum dot 1, in a regime where the exchange values to its nearest neighbors are much smaller than the Zeeman energy differences ($J_{12}, J_{15} \ll \Delta E_{15}, \Delta E_{12}$), favoring localization. In Fig.~\ref{fig:swaps}h, we tune the exchange couplings such that $J_{12} \approx J_{56} \approx 100$ MHz, more than an order of magnitude larger than the average Zeeman energy differences,  and  $J_{15} \approx J_{26} \approx 25$ MHz. We recognize two distinct oscillation frequencies: fast oscillations within each double dot pair, and a slow transfer of the spin excitation from the top to the bottom channel. This topology corresponds to two weakly coupled pairs of spins, where for the evolution time of \SI{150}{ns}, the Zeeman energy disorder does not play a role (see simulations of Fig. \ref{fig:weakly coupled}). We observe a decay of the signal after \SI{100}{ns}, consistent with a reduced coherence time at high exchange interaction values.

We furthermore explore the interesting regime in between the two aforementioned cases, where the exchange interactions and the Zeeman energy differences are similar (Figs. \ref{fig:swaps}i and j). For Fig. \ref{fig:swaps}i, using a previously reported method to equalize exchange interactions in a four-spin array \cite{van_diepen_quantum_2021,wang_probing_2023}, we tune all $J_{ij} \sim $ \SI{21}{MHz}, which is about 3.5 times larger than the average disorder in Zeeman energies. We observe an oscillation pattern between quantum dots 2 and 5, as expected for a ring with homogeneous exchanges, where the initial excitation only constructively interferes on the opposite site of the array. However, we observe an additional beating pattern, which we can attribute to the underlying Zeeman energy disorder (see simulations in Figs. \ref{fig:spin_ring_hom} and \ref{fig:spin_ring_sim}). Therefore, the observed dynamics is a consequence of the competition of both energy scales. Additionally, we leverage on our largely independent barrier control \cite{jirovec_exchange_2025} to explore the effect of locally introducing disorder in the exchange interactions. We find an exchange voltage configuration which seems to compensate for the underlying Zeeman energy disorder, yielding a regular oscillation pattern between dots 2 and 5 (Fig. \ref{fig:swaps}j). We again validate this with simulations (Figs. \ref{fig:spin_ring} and \ref{fig:spin_ring_sim}), from which we extract $J_{12} = $\SI{11.5}{MHz}, $J_{56} = $\SI{27.0}{MHz}, $J_{15} = $\SI{20.0}{MHz} and $J_{26} = $\SI{15.5}{MHz}. We further show our independent tunability of exchange interactions explicitly in Fig. \ref{fig:spin_ring_2D} by looking at the dependence of the oscillations on barrier gate voltages, which is in good agreement with simulations.

Looking closely at Figs.~\ref{fig:swaps}d,e, we observe that the oscillations as a function of SWAP time are slightly offset in phase compared to what one would ideally expect. Additionally, the oscillation frequency is below its steady value for short times $t_0$ in Figs.~\ref{fig:swaps}b,d,e. We attribute the former mainly to the use of 15 ns ramp times on the barrier pulses, and the latter to gate voltage pulse distortions.

Additionally, instead of changing the initialization ramp times to select which spin(s) to initialize in the excited state, one can also utilize the SWAP gates to move the initial excitation across the array before the analog evolution. We show this in Fig. \ref{fig:swaps}l (with Fig. \ref{fig:swaps}k showing the corresponding quantum circuit). For this experiment, we first prepare the full 2$\times$4 array in the state $\ket{2} = \ket{\downarrow_1\uparrow_2\downarrow_3...\downarrow_8}$ in all cases, and subsequently perform selected SWAP operations to take the excitation from dot 2 to any of the seven other dots. Finally, we read out pairs 3-4 and 5-6 followed by readout of pairs 1-2 and 7-8. Without (with) the readout SWAPs, the readout returns the probability of finding the excitation in dots 1, 4, 6 and 8 (2, 3, 5 and 7). As seen in the figure, we can successfully prepare any single-magnon product state with this method, with typical SWAP times below \SI{10}{ns}, and detect in which dot the magnon was initialized in two repetitions of the protocol. We note that, in this case, the contrast for the readout of dots 7 and 8 is quite low, which we attribute to a low readout visibility at the time of this measurement.

Note that these methods are not only limited to initializing single excitations but also allow to prepare up to four spin-up excitations, corresponding to a Néel state (or permutations thereof). This is demonstrated in Fig. \ref{fig:8spins_Neel}. Furthermore, we could append an intermediate-speed ramp back to (1,1) after PSB readout, such that the post-measurement state coincides with the pre-measurement state. Applying a SWAP operation and performing another PSB readout would allow us to record both $\uparrow\downarrow$ and $\downarrow\uparrow$ in a single execution. Ramping back semi-adiabatically once more and reading out again but now adiabatically, would give access to all four probabilities for each pair, and thus all eight single-spin probabilities, in every run, similar in spirit to a recent demonstration on two spins~\cite{nurizzo_complete_2023}.

\section{Quantum walk with singlet-triplet encoding}\label{sec:S-T$^{-}$_triplons}

\begin{figure*}
    \centering
    \includegraphics[width=\textwidth]{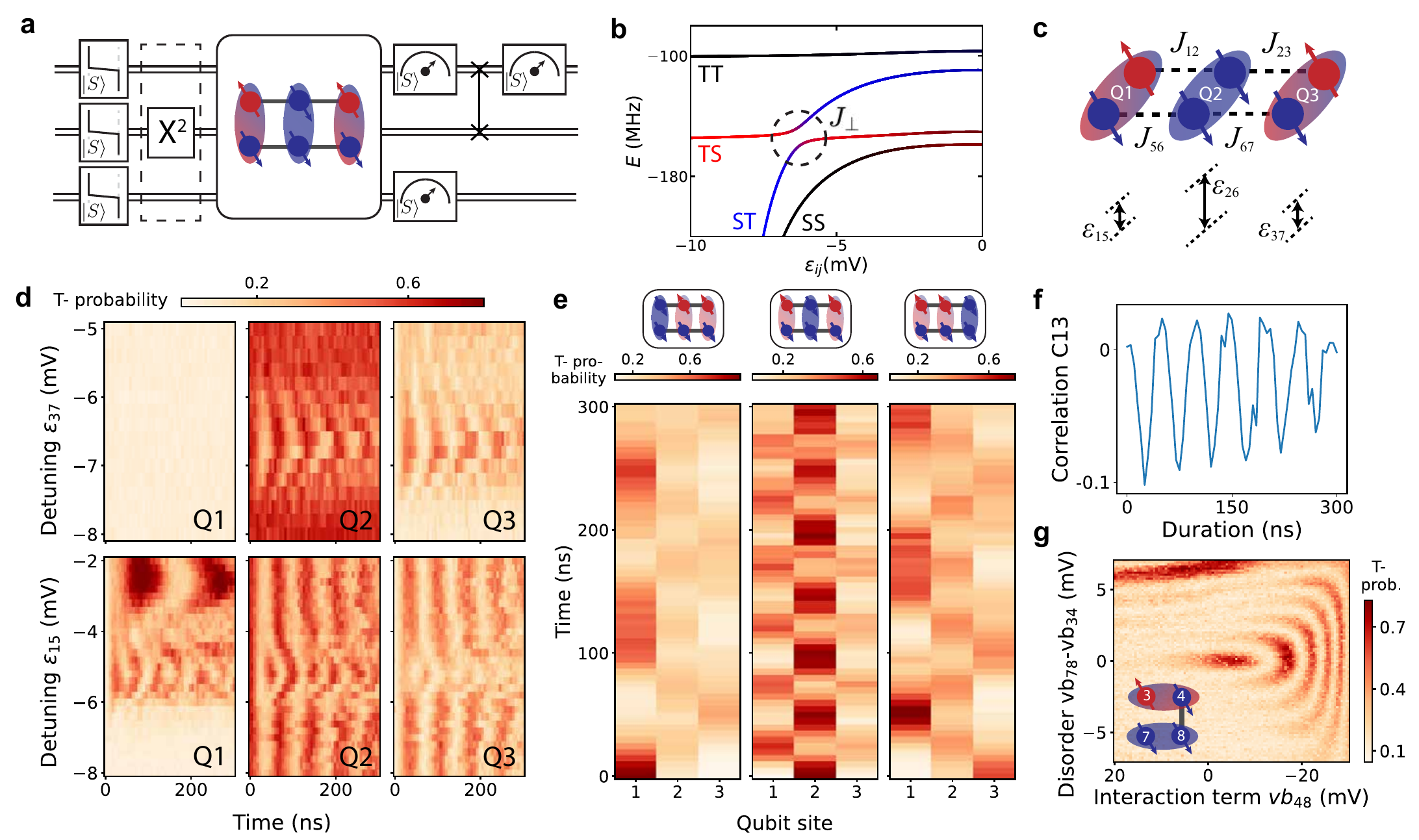}
    \caption{Triplon propagation in a dimerized quantum dot array. (a) Quantum circuit showing the state preparation and readout as well the analog evolution of a single triplon state. (b) Energy diagram for two coupled S-T$^{-}$ qubits, showing the four lowest energy branches as a function of detuning within one of the dot pairs. At a finite detuning value, an avoided crossing between the $ST^-$ and $T^-S$ branches occurs, the size of which is given by the average exchange coupling between both qubits. (c) Schematics of the detuning configuration at the start of the calibration procedure. (d) Two-step calibration of the homogeneous exchange point for three coupled singlet-triplet pairs. The measured $T^{-}$ probability is shown for each of the three qubits (as indicated) as a function of the detuning  $\varepsilon_{37}$ (top three panels) and  $\varepsilon_{15}$ (bottom three panels), and as a function of the analog evolution time. A triplon, initially confined in Q2, can propagate to Q3 (top) and through the full three-site system (bottom) at the respective homogeneous conditions (see main text). (e) Three-site quantum walk of a single triplon, initialized on Q1, Q2 or Q3, respectively, at the homogeneous condition. The $T^-$ probability is plotted for each of the three sites as a function of the analog evolution time. (f) Spin-spin correlation C$_{13}$ for the middle quantum walk of panel (e). (g) Two-site phase diagram as a function of interactions (vb$_{48}$, tuning $J_\parallel$ in Eq. \ref{eq:stham}) and disorder (vb$_{78}$ and vb$_{34}$, controlling $\bar{E_z} - J_\perp$). After initializing a state $\ket{S T^{-}}$ and allowing it to evolve for a fixed duration of \SI{200}{ns}, we retrieve the top site $T^{-}$ probability for different values of vb$_{48}$ and vb$_{78}$-vb$_{34}$.}
    \label{fig:qwalks_st}
\end{figure*}

We next turn to methods for exploring the propagation of two-spin excitations, also known as triplons \cite{dagotto_superconductivity_1992,giamarchi_quantum_2010,nawa_triplon_2019}, in the 2$\times$4 quantum dot ladder.  
The two-spin excitation is here given by the lowest-energy spin triplet state, T$^-$, with two-spin singlets along the rungs of the ladder forming the ground state. Singlet-triplet physics of a pair of quantum dots and qubit encodings in the singlet-triplet  basis have been widely explored in the literature \cite{jirovec_singlet-triplet_2021,jirovec_dynamics_2022,zhang_universal_2024}. 
Encoding a qubit in the S-T$^{-}$ subspace has the advantage that even at low magnetic fields, fast, high-fidelity and baseband single-qubit operations are available through rotations at the S-T$^{-}$ anticrossing \cite{zhang_universal_2024}, which at the \SI{10}{mT} in-plane field used for this experiment are as fast as \SI{10}{MHz}. Coupling two singlet-triplet qubits together results in an effective spin-spin Hamiltonian (see Eq. \ref{eq:stham}) where, in the language of propagating spin excitations, both the interaction and the disorder terms are gate-voltage tunable. Furthermore, this encoding allows for direct site-resolved readout via PSB. All this comes at the expense of halving the number of available sites, and we will operate the device in an effective 1D configuration.

Fig. \ref{fig:qwalks_st}a shows a circuit diagram depicting initialization, propagation and readout of three interacting singlet-triplet qubits. The initialized state $\ket{S_1T_2S_3}$ corresponds to a single triplon in the center of the array. After the analog evolution, successive PSB readouts (including a SWAP operation for spin transfer) enable the retrieval of all three singlet-triplet probabilities in every single run, in principle allowing us not only to extract the individual single-qubit probabilities, but also all spin-spin correlation terms $C_{ij} = P_{i,j}^T - P_i^TP_j^T$ \cite{fukuhara_microscopic_2013}, where $P_i^T$ is the probability of finding a triplon in site $i$, and $P_{i,j}^T$ is the joint probability of finding triplons in sites $i$ and $j$.

To observe magnon propagation without disorder, the terms $\propto S_{z,i}$ in Eq.~\ref{eq:stham} should be made equal for all pairs. This effectively requires satisfying the condition of Eq.~\ref{eq:stham_cond} for all qubit pairs $(i,j)$, which we call the homogeneous condition. For two qubits, this reduces to $\bar{E}_{z,i} - J_i^\perp = \bar{E}_{z,j} - J_j^\perp$, corresponding to the avoided crossing of Fig.~\ref{fig:qwalks_st}b, where we depict the energy levels of the S-T$^{-}$ subspace of two qubits as a function of the interdot detuning within each pair, which we use to control the respective exchanges.  At this avoided crossing, coherent $\ket{ST^{-}}\leftrightarrow\ket{T^{-}S}$ oscillations can be induced, with a frequency given by the average $J_{ij}^{\parallel}$ between the pairs~\cite{zhang_universal_2024}.

To tune to the homogeneous condition and enable magnon propagation through the array, we resort to an iterative method, illustrated in Fig.~\ref{fig:qwalks_st}c and d. First, we start with a detuning configuration away from the symmetry point $\varepsilon_{26} > \varepsilon_{15}, \varepsilon_{37}$, where the perpendicular exchange coupling corresponding to Q2 is the largest (Fig. \ref{fig:qwalks_st}c). Next, we scan the detuning $\varepsilon_{37}$, effectively increasing $J_3^\perp$. We observe coherent oscillations between Q2 and Q3 (top right panels of Fig.~\ref{fig:qwalks_st}d), which display a minimum frequency and a maximum visibility at the homogeneous condition. These are only visible in the Q2 and Q3 readout signals, as expected. Next, we fix the detuning values for Q2 and Q3 at this voltage point and scan $\varepsilon_{15}$. Away from the homogeneous condition, the speed of the Q2-Q3 oscillations remain unchanged, as expected. At the homogeneous condition point for all three qubits, a second set of oscillations is visible, which can now also be seen in the Q1 readout signal. This is a signature that the excitation that was first confined to pairs 2 and 3 can propagate all the way to pair 1. Again, at the minimum oscillation speed, all disorder (single-particle) terms are equal and the propagation speed is only given by the exact values of $J_{ij}^\parallel$. Note that the second set of oscillations visible in the bottom left panel of Fig. \ref{fig:qwalks_st}d corresponds to single-qubit rotations of Q1 (i.e. when $\bar{E}_{z,1} = J_1^\perp$), which occur well outside the homogeneous point, not affecting the magnon propagation. The data shown in Fig.~\ref{fig:qwalks_st}d correspond to tuning three singlet-triplet pairs, but this iterative method can be extended to larger arrays (see Fig. \ref{fig:ST_qwalk_4sites} for data on four pairs).

Fig. \ref{fig:qwalks_st}e shows quantum walk plots for three coupled pairs at the homogeneous condition. In each plot, a single triplon is initialized in different positions in the array. In this case, the readout fidelity of Q3 is significantly lower than that of Q1 and Q2, which explains the difference in oscillation amplitude between the columns. Simulating the quantum walk data with Eq. \ref{eq:stham} yields an excellent agreement (see Fig. \ref{fig:st_qwalk_sim}) and we extract $J^\parallel_{12} \approx J^\parallel_{23} =$ \SI{14}{MHz}. Additionally, we can extract all correlations $C_{ij}$ as shown in Fig. \ref{fig:ST_corr}. As an example, Fig. \ref{fig:qwalks_st}f shows the correlation value $C_{13}$ for the middle plot as a function of time. $C_{13}$ is negative at the point where the triplon is delocalized between sites 1 and 3: if it is measured to be in site 1, it will not appear in site 3, and vice versa. This is consistent with the expected generation of entanglement between Q1 and Q3, though measurements in additional bases would be needed to confirm the presence of entanglement~\cite{zhang_universal_2024}.

Finally, we explicitly demonstrate the tunability of interactions and disorder in Fig. \ref{fig:qwalks_st}g. For a two-site system (as depicted on the figure inset), we initialize $\ket{S T^{-}}$ and evolve for a fixed duration of \SI{200}{ns}, before retrieving the triplet probability of the top channel. We do so as a function of virtual barrier voltage vb$_{48}$, which controls the coupling $J_{48}$ between the two sites, and barriers vb$_{34}$ and vb$_{78}$ which control $J_{34}$ and $J_{78}$, respectively, and therefore the single-site disorder for the top and bottom sites. At the relative value vb$_{78}-$vb$_{34} = 0$, the system is at the homogeneous condition with no disorder between the two sites, and oscillations are observed for a wide range of values of vb$_{48}$. As vb$_{78}$ and vb$_{34}$ are asymmetrically scanned away from zero, the single-site disorder between the two sites increases, leading to localization of the triplon for small $J_{48}$ (positive vb$_{48}$). Only when $J_{48}$ is large enough to overcome disorder (negative vb$_{48}$), the triplon propagates. Note that there is an extra line of high triplet probability on the top right corner of the plot, consistent with the S-T$^-$ anticrossing of quantum dot pair 3-4. Overall, this plot can be viewed as a two-site phase diagram, and highlights the independent tunability of the interaction and single-site disorder terms.

\section{Discussion and outlook} \label{sec:disc}

In this work, we have observed the dynamics of spin excitations as they propagate through an array of exchange-coupled quantum dots. We expand upon the methodologies for initialization, manipulation and readout of germanium quantum dot spin qubits, while also building on analog simulation techniques for semiconductor quantum dot arrays. This analog-digital framework has allowed us to initialize, control and track the propagation of excitations in an extended quantum dot array, while maintaining a high degree of control over individual Hamiltonian parameters. This highlights the power offered by the combination of analog and digital techniques for quantum simulations with quantum dots. By exploring two different types of spin excitations, magnons and triplons, we present the advantages, operation techniques and challenges of both approaches.

We note that in the present work we neglected the tensorial components of the exchange interaction terms, as similarly done in a related work~\cite{wang_probing_2023}. Since we are able to model the quantum walk plots with an isotropic model, we speculate that this anisotropy is either small for our particular device, or that we successfully avoid unwanted rotations on any spin-orbit induced anticrossing in the full energy spectrum during the analog evolution. Although outside the scope of this work, and since these effects might become more relevant for larger arrays or more complex geometries, we encourage further research to quantitatively and systematically model and measure the $J$-tensor of coupled germanium quantum dots, as done in the literature for hole spins in silicon dots~\cite{geyer_anisotropic_2024}. In this regard, it would be worthwhile to investigate spin dynamics at different magnetic field strengths and orientations, as the spin Hamiltonian parameters, including spin-orbit terms, have an explicit dependence on the magnetic field orientation \cite{jirovec_dynamics_2022,geyer_anisotropic_2024,saez-mollejo_exchange_2025,hendrickx_sweet-spot_2024}. 

Interestingly, by coupling all eight spins and observing the evolution of single-spin probabilities of highly excited states such as the Néel state, one could already explore the thermalization properties of the system under the respective spin Hamiltonians as a function of disorder, spin-spin interaction strengths and topologies. While such measurements could yield a first indication of the many-body localized or ergodic phases, one could additionally resort to further witnesses such as excited state spectroscopy and energy level distribution~\cite{jurcevic_spectroscopy_2015,roushan_spectroscopic_2017,serbyn_spectral_2016,oganesyan_localization_2007} among others, to further probe the nature of the underlying Hamiltonians. 

Such studies will cross the many-body threshold as the quantum dot arrays are scaled up to larger sizes while maintaining a high level of control, high initialization and readout fidelities and state coherence. We envision that these techniques can be readily applied to extended (quasi one-dimensional) quantum dot ladders, and, with more effort, to two-dimensional and even three-dimensional~\cite{ivlev_coupled_2024} arrays. Still, there are several challenges to tackle moving in this direction. Even in the present work, the precise and simultaneous tuning of all interactions within the array was slowed down by the fact that any single gate voltage affects the on-site energy and tunnel coupling of neighboring sites. We have nonetheless been able to dial in the Hamiltonian parameters largely independently through the use of virtualization techniques~\cite{qiao_coherent_2020,van_diepen_quantum_2021,wang_probing_2023} and have expanded on these in a recent, related work \cite{jirovec_exchange_2025}, enabling the present experiments. Although the present results still remain in the few-body regime and can be efficiently benchmarked with classical simulations based on exact diagonalization, we achieve significant milestones that can potentially be scaled up to larger systems, likely with the aid of automated tuning routines~\cite{rao_modular_2025,hsiao_efficient_2020}. All in all, this work lays the groundwork for the exploration of the rich phase diagram and excitation dynamics in extended single-spin and dimer lattices.

\section*{Acknowledgments}

We acknowledge the contribution of A. Sammak to the development of the Ge/SiGe heterostructure. We thank M. Rimbach-Russ, S. Bosco, V. John, F. Borsoi, C.-A. Wang and other members of the Vandersypen, Veldhorst, Scappucci, Rimbach-Russ and Bosco groups for stimulating discussions. We acknowledge technical support by O. Benningshof, J. D. Mensingh, T. Orton, R. Schouten, R. Vermeulen, R. Birnholtz, E. Van der Wiel, B. Otto and D. Brinkman. This work was funded by an Advanced Grant from the European Research Council (ERC) under the European Union’s Horizon 2020 research (882848).

\section*{Data availability}
The raw measurement data and the analysis supporting the
findings of this work are available on a Zenodo repository (https://doi.org/10.5281/zenodo.15607472).

\bibliography{references}

\clearpage

\widetext
\begin{center}
\textbf{\large Supplemental Materials: Site-resolved magnon and triplon dynamics on a quantum dot spin ladder}
\end{center}

\setcounter{equation}{0}
\setcounter{figure}{0}
\setcounter{table}{0}
\setcounter{page}{1}
\makeatletter
\renewcommand{\theequation}{S\arabic{equation}}
\renewcommand{\thefigure}{S\arabic{figure}}
\renewcommand{\thetable}{S\arabic{table}}

\section{Device fabrication and experiment setup}

The devices used in this work were fabricated on a Ge/Si$_{0.2}$Ge$_{0.8}$ heterostructure grown on a silicon wafer, with the quantum well buried \SI{55}{nm} below the surface. Details about the heterostructure and the growth process can be found in Ref. \cite{lodari_low_2021}. 

\begin{figure*}[h]
    \centering
    \includegraphics[width=0.95\textwidth]{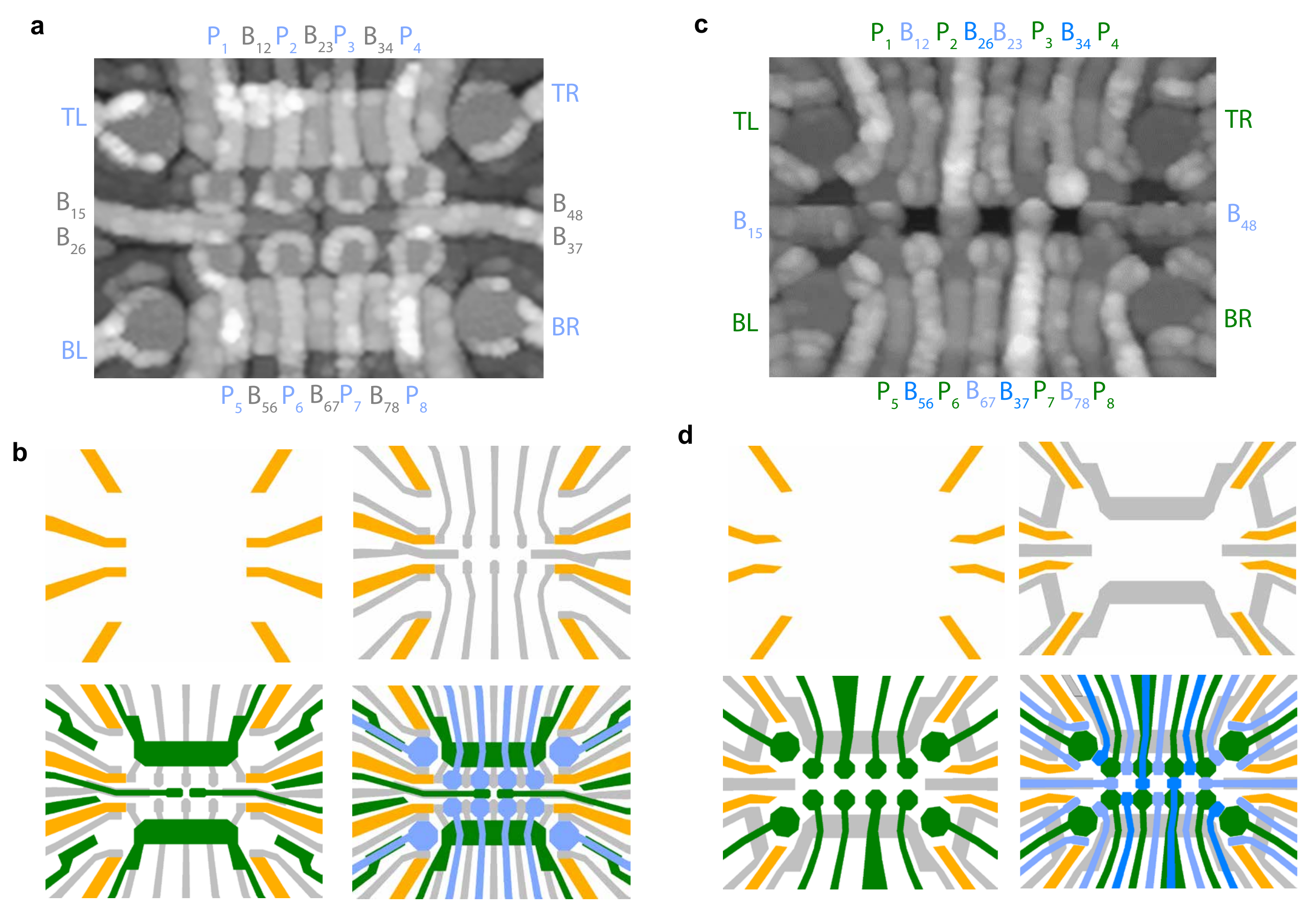}
    \caption{(a) AFM of 2$\times$4 device used for the magnon propagation experiments. (b) Corresponding multi-layer gate layout. The four panels display the four different fabrication layers: Ohmic contacts, barrier gates, screening gates and plunger gates. Between each fabrication step, an Al$_2$O$_3$ ALD layer is deposited, preventing contact between different fabrication layers. (c) AFM of 2$\times$4 device used for the triplon propagation experiments. (d) Corresponding multi-layer gate layout. The four panels display the four different fabrication layers: Ohmic contacts, screening gates, plunger gates and barrier gates. Note the difference in ordering with respect to panel (b). Between each fabrication step, an Al$_2$O$_3$ ALD layer is deposited, preventing contact between different fabrication layers. The barrier layer was deposited in two steps to avoid unwanted shorting of the gates.}
    \label{fig:device_layout}
\end{figure*}

Fig. \ref{fig:device_layout} shows two atomic-force microscopy images of the two devices used in this work. Device 1 (Fig. \ref{fig:device_layout}a) was used for the magnon propagation experiment, with device 2 (Fig. \ref{fig:device_layout}b) being used to measure triplon propagation. As a first fabrication step, Pt (device 1) or Al (device 2) Ohmics are defined using electron beam lithography and subsequently annealed so that the metal diffuses into the heterostructure to directly contact the quantum well. Subsequently, three layers of Ti/Pd gate electrodes are patterned to electrostatically define the potential landscape, with thicknesses of 3/\SI{17}{nm}, 3/\SI{27}{nm} and 3/\SI{37}{nm}, respectively. The three layers correspond to plunger gates, which control the electrochemical potential of the quantum dots; barrier gates, to control the tunnel coupling and exchange interaction between nearest neighbors; and screening gates to screen unwanted accumulation of charges near the device's active area. The fabrication order of the gate electrodes is barriers-screening-plungers for device 1 (Fig. \ref{fig:device_layout}c) and screening-plungers-barriers for device 2 (Fig. \ref{fig:device_layout}d). All gate layers are separated from each other with Al$_2$O$_3$ grown by atomic layer deposition (ALD), each of them \SI{5}{nm} thick. 

The experiments are carried out in Oxford Triton (device 1) and Bluefors XLD (device 2) dilution refrigerators with base temperatures of around \SI{10}{mK}. Each gate electrode is connected with DC lines to the ports of home-built serial peripheral interface (SPI) DAC modules to individually set DC voltages. In addition, all plunger and barrier gates are connected to a Keysight M3202A arbitrary waveform generator (AWG) for baseband pulse control. Both DC and AC signals are combined on the printed circuit board (on which the sample is glued and bonded) using bias tees with resistor-capacitor time constants of \SI{100}{ms}, orders of magnitude higher than the duration of each single experimental shot. Between shots, a compensation pulse is applied to all AC gates whose area is equal and opposite to the accumulated DC offset during the measurement to mitigate bias-tee charging. AC lines are attenuated and thermally anchored to each dilution refrigerator plate using cryogenic attenuators. Additionally, these lines are filtered using common-mode ferrite chokes at room temperature to filter out low-frequency noise. The high-frequency noise on the DC lines is filtered using resistor-capacitor filters and copper-powder filters at the mixing chamber plate. 

Charge sensors are used to read out the charge state of the devices as a result of their capacitive coupling to neighboring dots. The state of the sensors themselves is read out using radiofrequency (RF) reflectometry via one of the two Ohmic contacts, which is bonded to an off-chip NbTiN inductor forming a inductor-capacitor (LC) tank circuit. Every inductor is fabricated to have a different inductance, which results in a different resonance frequency for each tank circuit. This allows for simultaneous readout of all sensors by multiplexing all Ohmics to a single RF line, and modulation and demodulation measurements using custom-built SPI RF generators and in-phase and quadrature (IQ) demodulators. 

Note that, while the loading of charges into the quantum dot array happens via the sensing dot for device 2, which requires the sensor to be tunnel-coupled to its neighboring quantum dot, loading happens directly from the Ohmics in device 1. In this case, half of the Ohmics are shared between sensing dot and quantum dot (see device schematics in Figs. \ref{fig:device_layout}c and d). This comes at the expense of increasing the sensor-to-dot distance by about \SI{30}{\%}, which does not significantly affect the sensing capabilities, but which makes the independent tunability of sensing dot and quantum dots easier.

\clearpage

\section{Charge stability diagram, initialization and readout ramps}

\begin{figure*}[h]
    \centering
    \includegraphics[width=0.95\textwidth]{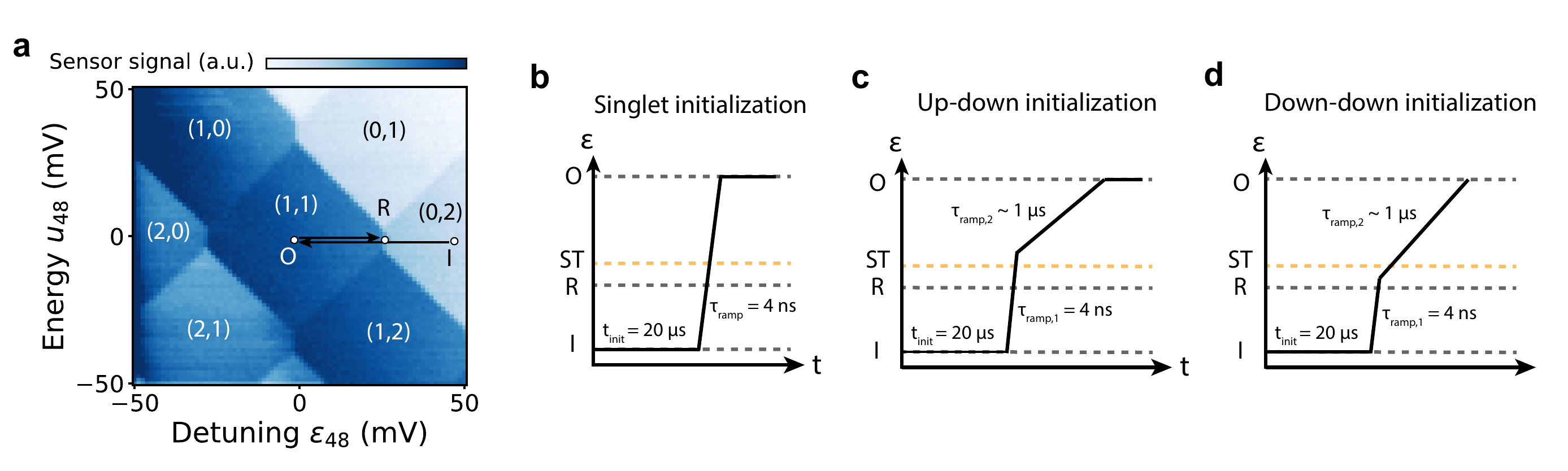}
    \caption{Schematics of pulsed initialization. (a) Charge stability diagram of quantum dot pair 4-8. Along the x-axis, we scan the detuning $\varepsilon_{48} = \frac{1}{2}(vp_8 - vp_4)$, and the global chemical potential $u_{48} = vp_8 + vp_4$ is scanned along the y-axis. For each combination of gate voltages, we record the sensor signal. The numbers $(n_1,n_2)$ represent the double dot charge occupation. The three voltage points I (initialization), O (operation) and R (readout) lie along the detuning axis and are calibrated for each quantum dot pair.  (b) Pulse scheme to initialize the state $\ket{S(1,1)}$. After waiting \SI{20}{\micro s} at point I to initialize the state $\ket{S(2,0)}$, we pulse to the operation point with a ramp speed of \SI{4}{ns}, which is diabatic and conserves the total spin state. (b) Pulse scheme to initialize the state $\ket{\uparrow\downarrow}$. After waiting \SI{20}{\micro s} at point I, we first pulse fast (\SI{4}{ns}) just over the ST$^-$ anticrossing (shown as a yellow dashed line), conserving the spin singlet state. Subsequently, a slow ($\sim$ \SI{1}{\micro s} ramp) is used, which adiabatically maps the singlet into the $\ket{\uparrow\downarrow}$ state. (c) Pulse scheme to initialize the state $\ket{\downarrow\downarrow}$. After waiting \SI{20}{\micro s} at point I, we first pulse fast (\SI{4}{ns}) just before the ST$^-$ anticrossing. Subsequently, a slow ($\sim$ \SI{1}{\micro s} ramp) is used, which adiabatically maps the singlet into the $\ket{\downarrow\downarrow}$ state.}
    \label{fig:csd_init}
\end{figure*}

In Fig. \ref{fig:csd_init}, we schematically represent the pulse scheme used for pairwise initialization throughout this work. To navigate the charge state of the system and correctly calibrate the detuning axes, we first record the charge stability diagrams (Fig. \ref{fig:csd_init}a) for each quantum dot pair. Most detuning pulses used in this work rely on three different voltage points: ``I", the initialization point deep in the (0,2) charge state, where a singlet is initialized after a typical waiting time of \SI{20}{\micro s}; ``O", the operation point in the center of the (1,1) charge configuration; and ``R", the readout point where Pauli spin blockade is performed. As outlined in the main text, we use three different ramp configurations between ``I" and ``O" to initialized three different states: S(1,1) (\ref{fig:csd_init}b), $\ket{\uparrow\downarrow}$ (\ref{fig:csd_init}c) and $\ket{\downarrow\downarrow}$ (\ref{fig:csd_init}d). By choosing either fast or slow ramps, we ensure to be either adiabatic or diabatic with respect to the ST$^-$ anticrossing (which is represented as a yellow dashed line at the corresponding detuning value) and the Zeeman energy difference $\Delta E_z$. More information is outlined in the main text. Figs. \ref{fig:csd_init}b-d correspond to the schematics of Figs. \ref{fig:initread}b-d.

\clearpage

\section{Two-spin dynamics for all four pairs}

\begin{figure*}[h]
    \centering
    \includegraphics[width=0.95\textwidth]{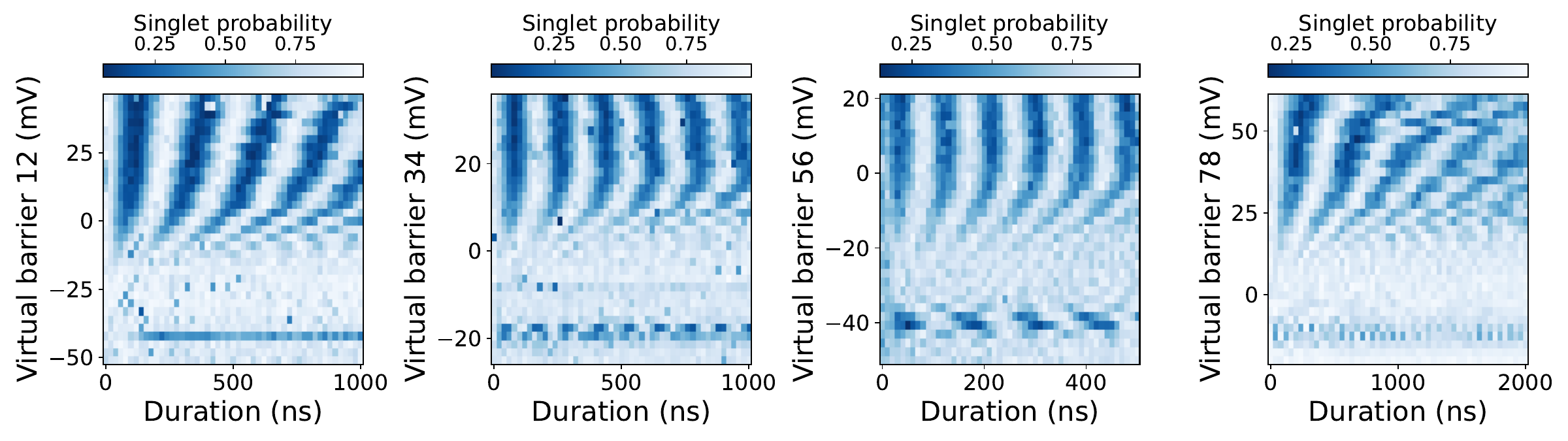}
    \caption{Dynamics of an initial state $\ket{S}$ as a function of interaction strength for each pair of quantum dots in the array. At positive barrier voltages, we observe ST$^0$ oscillations, corresponding to the regime of low exchange. At more negative barrier voltages (high exchange interaction), the singlet is an eigenstate and does not oscillate. Each of the four pairs shows oscillations at the ST$^-$ anticrossing at $J = \bar{E}_z$.}
    \label{fig:st_all}
\end{figure*}

\begin{figure*}[h]
    \centering
    \includegraphics[width=0.95\textwidth]{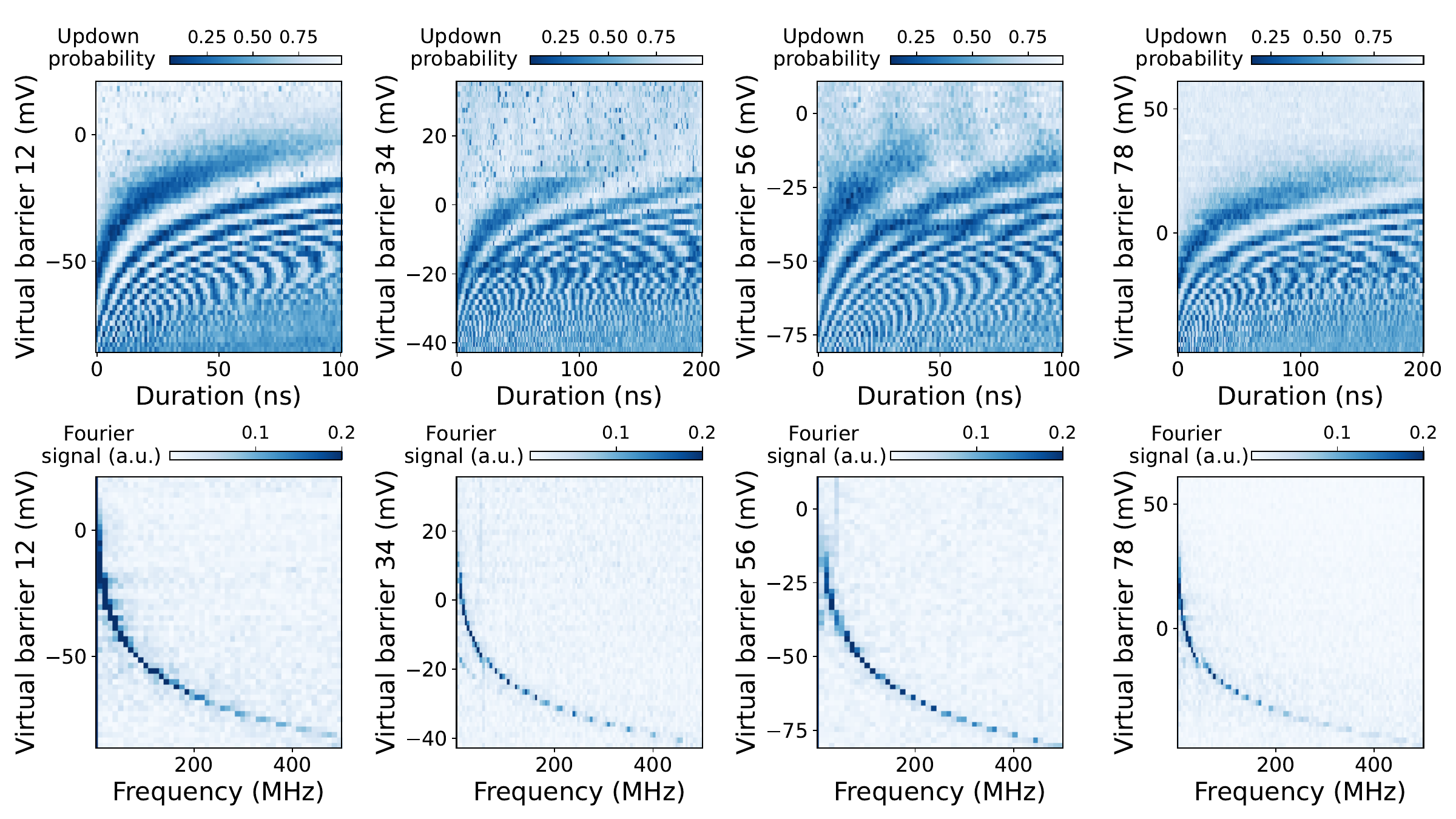}
    \caption{Top: Dynamics of an initial state $\ket{\uparrow\downarrow}$ as a function of interaction strength for each pair of quantum dots in the array. At positive barrier voltages, $\ket{\uparrow\downarrow}$ is an eigenstate and does not oscillate. At more negative barrier voltages (high exchange interaction), we observe SWAP oscillations. Each of the four pairs shows oscillations at the ST$^-$ anticrossing at $J = \bar{E}_z$. Bottom: corresponding Fourier transforms of the signal, showing oscillation speeds reaching \SI{500}{MHz} for all pairs. These Fourier spectra are fitted to extract the exchange values reported in Fig. \ref{fig:initread}h.}
    \label{fig:magnon_all}
\end{figure*}

\clearpage

\section{Coherent SWAPs at high exchange for pair 34}

\begin{figure*}[h]
    \centering
    \includegraphics[width=0.75\textwidth]{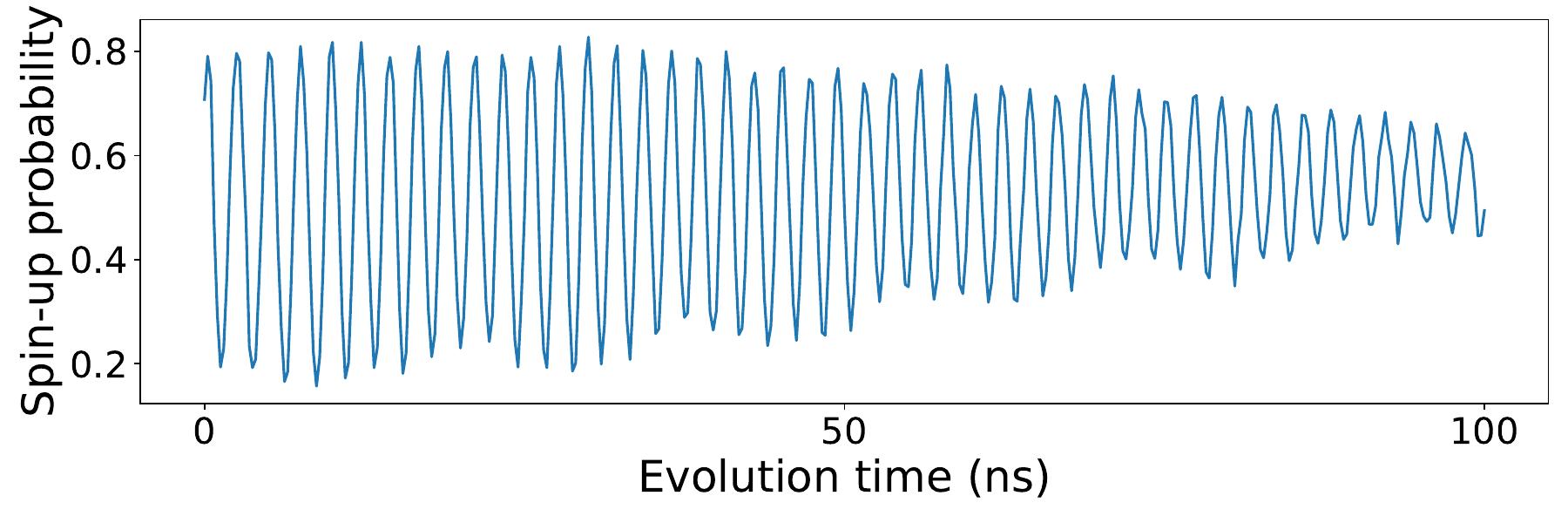}
    \caption{Fast exchange oscillations of an initial state $\ket{\uparrow\downarrow}$ on quantum dot pair 34. The obtained exchange speed is $J =$ \SI{460}{MHz}. Even at this high exchange, for this particular quantum dot pair, we retain a high quality factor, highlighting the potential operation of quantum dot devices at high exchange coupling values, as well as ultra-fast two-qubit gates in the regime where $J \gg \Delta E_z$.}
    \label{fig:Suppswaps_highj}
\end{figure*}
\clearpage

\section{Simulation of energy diagram and initialization fidelity}

\begin{figure*}[h]
    \centering
    \includegraphics[width=0.5\textwidth]{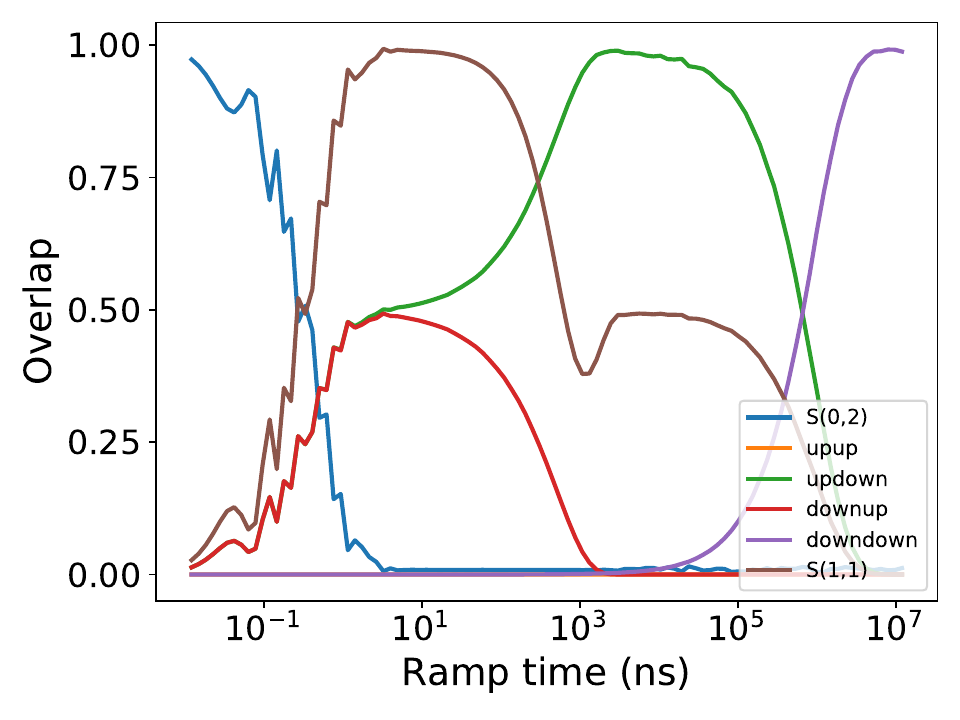}
    \caption{Simulation of initialization fidelity as a function of ramp time from the (2,0) to the (1,1) state. The simulation shows four distinct regimes. (a) S(0,2) initialization: the ramp time is diabatic with respect to tunnel coupling. (b) S(1,1) initialization: ramping is adiabatic with respect to tunnel coupling, but diabatic with respect to all spin energy scales. (c) $\ket{\downarrow\uparrow}$ initialization: ramping is diabatic with respect to $\Delta_{\text{ST}}$ and adiabatic otherwise. (d) $\ket{\downarrow\downarrow}$ initialization: ramping is adiabatic with respect to all energy scales. For this simulation, we include not only a ramp in detuning but also on tunnel coupling, consistent with experimental values. Simulation input parameters: $U = U_1 = U_2 = \SI{300}{GHz}$, $E_{z,1} = \SI{40}{MHz}$, $E_{z,2} = \SI{55}{MHz}$, $t_{\text{SO,x}} = t_{\text{SO,y}} = t_{\text{SO,z}}  = \SI{20}{MHz}$, $\varepsilon_0 = -2 U$, $\varepsilon_f =  0$,  $t_0 = \SI{20}{GHz}$, $t_f = \SI{0.2}{GHz}$. For this simulation, the ramp is chosen to be linear, while in experiments, we resort to composite ramps to increase initialization speed and fidelity.}
    \label{fig:init_sim}
\end{figure*}

The Hamiltonian to describe the lowest energy levels of a double quantum dot is given by \cite{jirovec_dynamics_2022}:

\begin{equation}
H=
\begin{pmatrix}
    U + \varepsilon & -t_{\text{so}}^x + it_{\text{so}}^y  &  t_0 - it_{\text{so}}^z  &  -t_0 - it_{\text{so}}^z  &  -t_{\text{so}}^y - it_{\text{so}}^x \\
    -t_{\text{so}}^y - it_{\text{so}}^x & \bar{E_z} & 0 & 0 & 0\\
    t_0 + it_{\text{so}}^z & 0 & \Delta E_z & 0 & 0\\
    -t_0 + it_{\text{so}}^z & 0 & 0 & -\Delta E_z & 0\\
    -t_{\text{so}}^y + it_{\text{so}}^x & 0 & 0 & 0 & -\bar{E_z}\\
\end{pmatrix},
\label{eq:FH_dd}
\end{equation}
\newline 
written in the basis $\{S(0,2), \ket{\downarrow\downarrow}, \ket{\downarrow\uparrow}, \ket{\uparrow\downarrow}, \ket{\uparrow\uparrow}\}$.

To simulate state initialization as a function of ramp time, we solve the time-dependent Schrödinger equation

\begin{equation}
    i \frac{\partial}{\partial t}\ket{\psi(t)} = H(t) \ket{\psi(t)}
\end{equation}

using QuTip. We introduce the time dependency in the detuning as $\varepsilon(t) = \varepsilon_0 + vt$ and $t_c(t) = t_0 - vt$, where $v$ is the ramp rate in GHz/ns. The final state can be simulated by finding the evolution operators

\begin{equation}
    U(t, t+\Delta t) = e^{-iH(t)\Delta t}
\end{equation}

for small $\Delta t$ values and sequentially calculating the state vectors at every time step. The results can be seen in Fig. \ref{fig:init_sim}. We observe four distinct regimes emerge, depending on the degree of (a-)diabaticity with respect to three energy scales: tunnel coupling $t_c$, Zeeman energy difference $\Delta E_z$ and spin-orbit anticrossing $\Delta_{ST}$. In the simulation, while using realistic experimental parameters, we make sure that these timescales are separable. In the experiment, if the spin-orbit anticrossing is too large or the difference in Zeeman energy too small, we resort to composite ramps as highlighted in the main text, which allow us to cross each anticrossing individually with the desired speed and ensure proper initialization of all relevant states.

\clearpage

\section{Simulation of oscillation patterns of Fig. \ref{fig:initread}}

\begin{figure*}[h]
    \centering
    \includegraphics[width=0.9\textwidth]{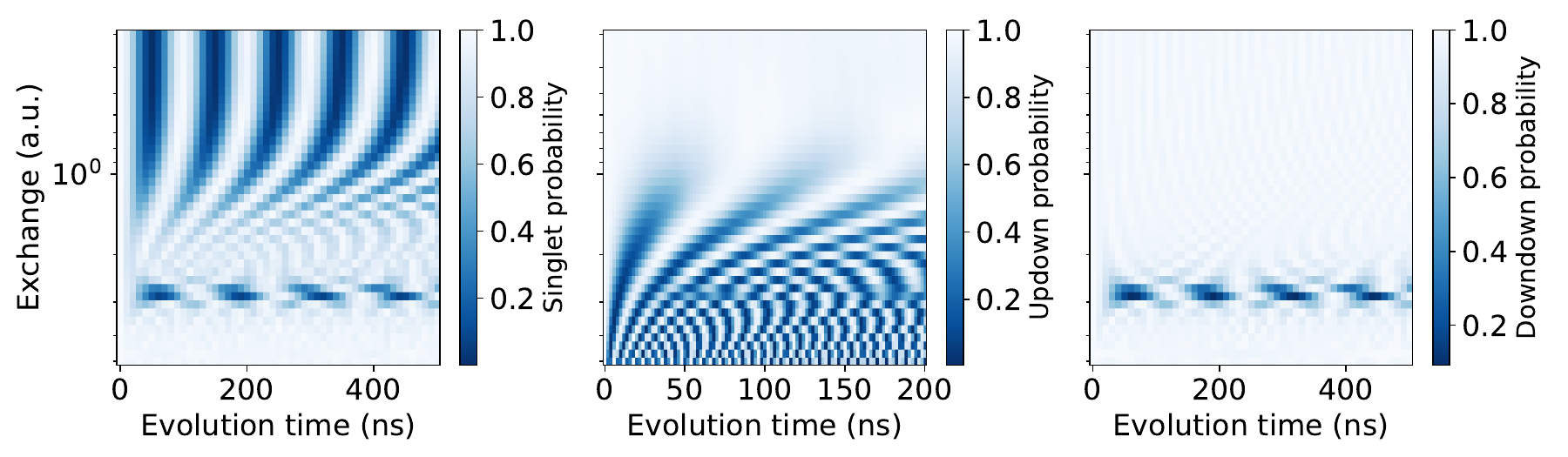}
    \caption{Simulation of oscillation patterns in Fig.\ref{fig:initread}b-d. The simulated results match with $E_{z,1} = \SI{50}{MHz}$ and $E_{z,2} = \SI{55}{MHz}$, slightly higher than the Zeeman energy values we find in Fig. \ref{fig:initread}h. A possible explanation is $g$-factor modulation through the barrier pulses, which also explains the strong bending of the S-T$^0$-oscillations for higher barrier voltages. For these simulations, we use the initial states that result from the ramped initialization simulations of Fig. \ref{fig:init_sim} for the ramp speeds with the highest overlap with the desired state.}
    \label{fig:osc_all_sim}
\end{figure*}

\clearpage

\section{EDSR spectra}

\begin{figure*}[h]
    \begin{floatrow}
    \ffigbox{%
    \centering
    \includegraphics[width=0.5\textwidth]{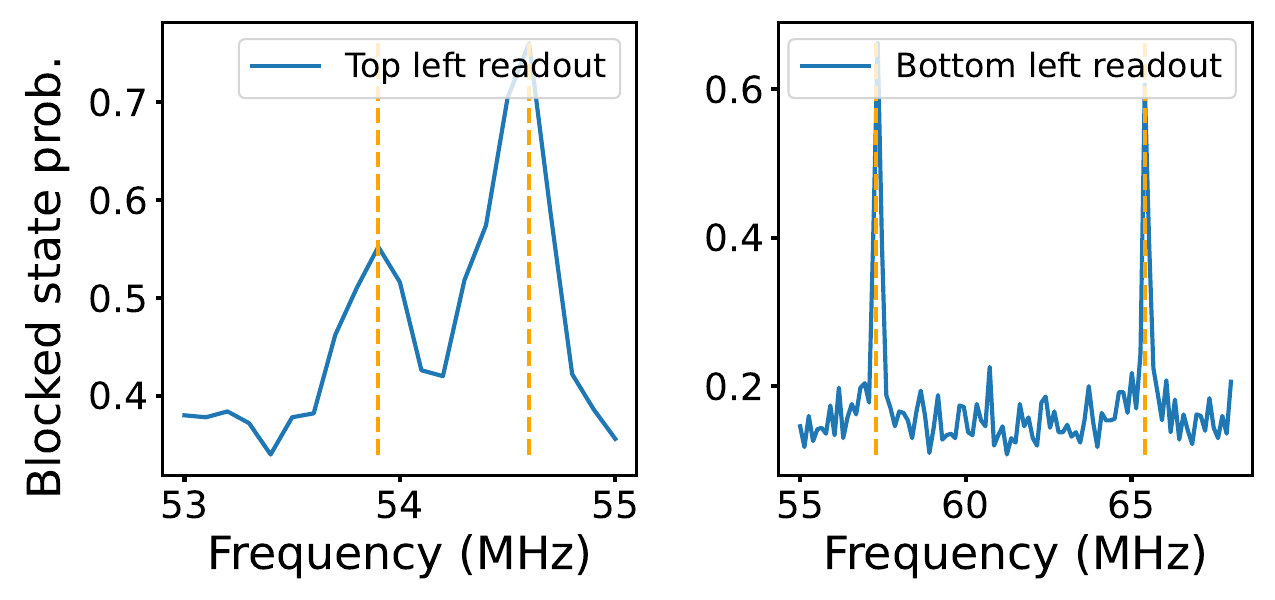}
    }{%
      \caption{EDSR spectra corresponding to the resonance frequencies of spins 1 and 2 (left panel) and 5 and 6 (right panel) for the magnon experiment, measured during the first cooldown. At this point, only the left half of the 2$\times$4 array was tuned, resulting in different resonance frequencies as compared to those reported in Fig. \ref{fig:initread}h.}%
      \label{fig:supp_edsr_spectra}
    }
    \capbtabbox{%
      \begin{tabular}{>{\centering\arraybackslash}m{2cm} >{\centering\arraybackslash}m{3cm} >{\centering\arraybackslash}m{3cm}}
\toprule
      Qubit number & Resonance frequency (first cooldown) & Resonance frequency (second cooldown) \\ \hline
      Q1 & 54.6 & 58.2\\
      Q2 & 53.9  & 62.8\\ 
      Q3 & -- & 58.4\\
      Q4 & -- & 51.8\\ 
      Q5 & 57.3 & 51.7\\
      Q6 & 65.4 & 42.4\\ 
      Q7 & -- & 48.8\\
      Q8 & -- & 44.2\\
      \end{tabular}
    }{%
      \caption{Summary of EDSR frequencies for Fig. \ref{fig:supp_edsr_spectra} (left table) and Fig. \ref{fig:initread}h, corresponding to the two different cooldowns and tuning regimes used for the data in this paper. The disorder landscape of the left table corresponds to the quantum walk plots of Fig. \ref{fig:swaps}g-j, while the right table corresponding  to all other plots of Figs. \ref{fig:initread} and \ref{fig:swaps}.}%
      \label{tab:edsr_spectra}
    }
    \end{floatrow}
\end{figure*}

\begin{figure*}[h]
    \centering
    \includegraphics[width=0.8\textwidth]{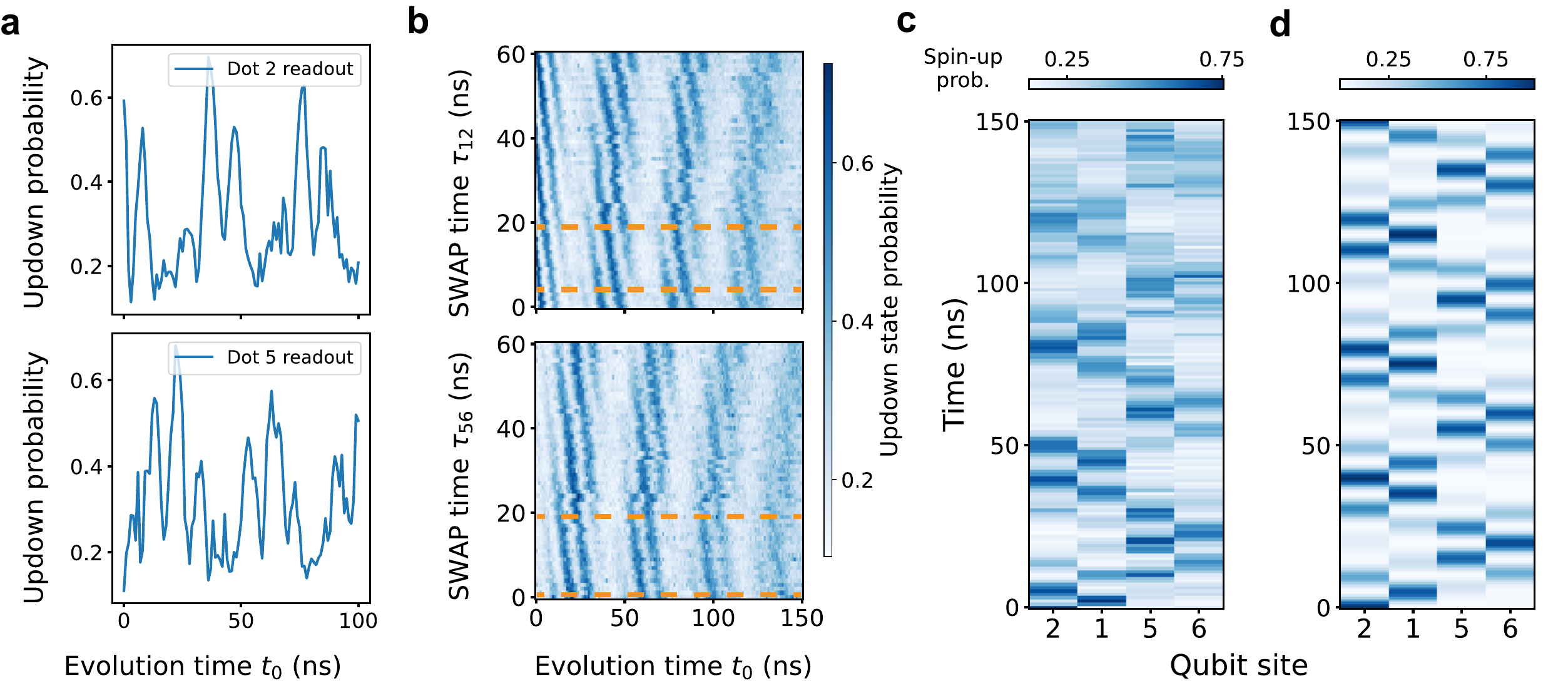}
    \caption{Measurements and simulation of the dynamics of a single excitation $\ket{2} = \ket{\downarrow_1\uparrow_2\downarrow_3\downarrow_4}$ in a 2$\times$2 array of weakly coupled pairs, corresponding to Fig. \ref{fig:swaps}h. (a) Unblocked state probability for pair 12 (top panel) and pair 56 (bottom panel) after a variable evolution time $t_0$. The readout was performed ramping with an intermediate speed to the PSB point, where the unblocked state is expected to be $\ket{\uparrow\downarrow}$ (see main text). For a single excitation in the array and the $g$-factor landscape at this time of tuning, this corresponds to the spin-up probabilities of dots 2 and 5, respectively. (b) Readout as a function of SWAP time $\tau_{12}$ (top panel) and $\tau_{56}$ (top panel). The dashed horizontal lines correspond to rotations by $\pi$ and $2\pi$ (top panel) or 0 and $\pi$ (bottom panel), which can be inferred by comparison to the measurements without the SWAP gate (panel (a)). The dashed lines then correspond to all four single-spin probabilities. (c) Site-resolved quantum walk, where the spin-up probabilities for each site are plotted as a function of evolution time. (d) Simulated quantum walk for this exchange coupling configuration, and using the Zeeman energy values of table \ref{tab:edsr_spectra} (left column), showing good agreement with the measured data. From the simulation, we extract $J_{12} = J_{56} =$ \SI{100}{MHz} and $J_{15} = J_{26} =$ \SI{25}{MHz}.}
    \label{fig:weakly coupled}
\end{figure*}

\begin{figure*}[h]
    \centering
    \includegraphics[width=0.8\textwidth]{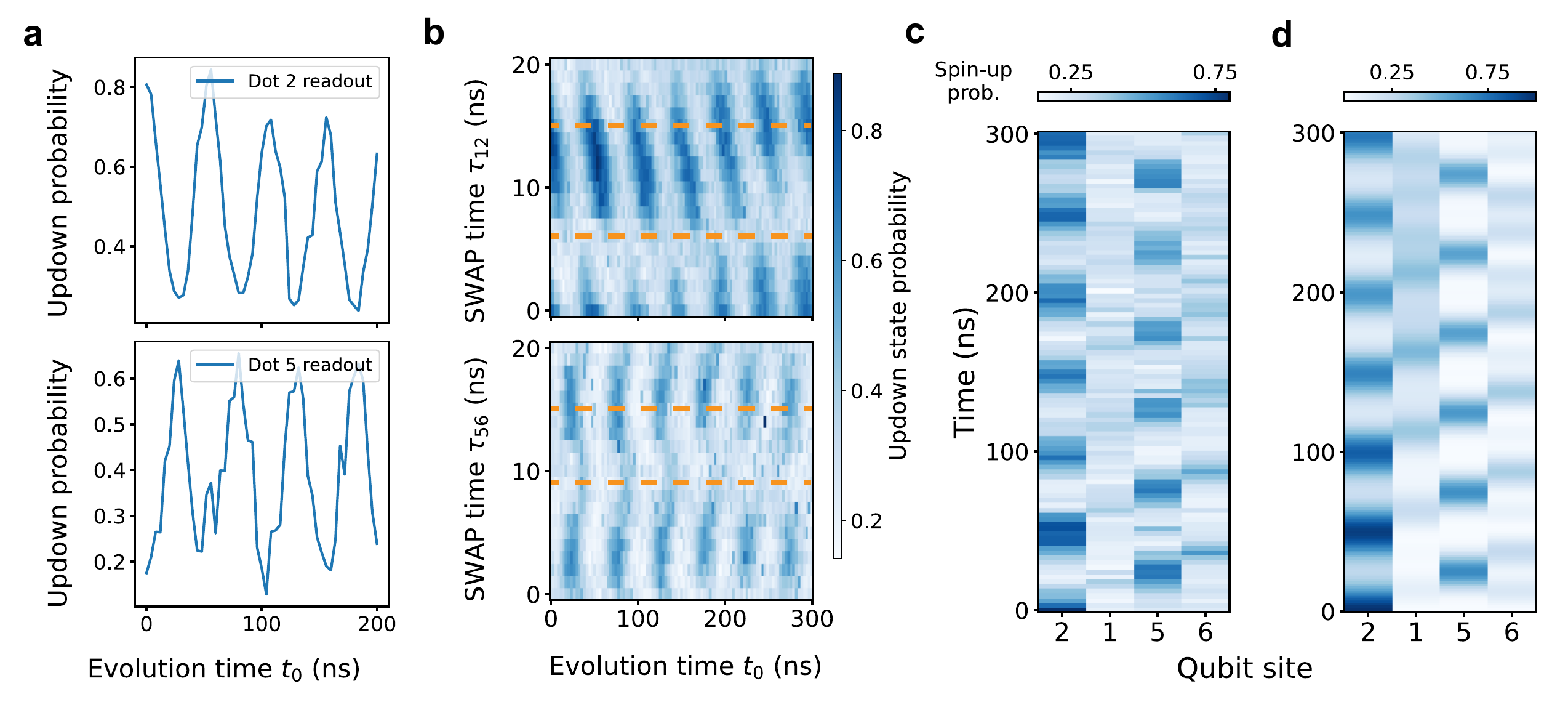}
    \caption{Measurements and simulation of the dynamics of a single excitation $\ket{2} = \ket{\downarrow_1\uparrow_2\downarrow_3\downarrow_4}$ in a 2$\times$2 array with all exchange couplings on, corresponding to Fig. \ref{fig:swaps}j. (a) Unblocked state probability for pair 12 (top panel) and pair 56 (bottom panel) after a variable evolution time $t_0$. The readout was performed ramping with an intermediate speed to the PSB point, where the unblocked state is expected to be $\ket{\uparrow\downarrow}$ (see main text). For a single excitation in the array and the $g$-factor landscape at this time of tuning, this corresponds to the spin-up probabilities of dots 2 and 5, respectively. (b) Readout as a function of SWAP time $\tau_{12}$ (top panel) and $\tau_{56}$ (top panel). The dashed horizontal lines correspond to rotations by $\pi$ and $2\pi$ (top panel) or $2\pi$ and $3\pi$ (bottom panel), which can be inferred by comparison to the measurements without the SWAP gate (panel (a)). The dashed lines then correspond to all four single-spin probabilities. (c) Site-resolved quantum walk, where the spin-up probabilities for each site are plotted as a function of evolution time. (d) Simulated quantum walk for an exchange coupling configuration $J_{12} =$ \SI{11.5}{MHz}, $J_{56} =$ \SI{27.0}{MHz}, $J_{15} =$ \SI{20.0}{MHz} and $J_{26} =$ \SI{15.5}{MHz}, and using the Zeeman energy values of table \ref{tab:edsr_spectra} (left column), showing good agreement with the measured data.}
    \label{fig:spin_ring}
\end{figure*}

\begin{figure*}[h]
    \centering
    \includegraphics[width=0.6\textwidth]{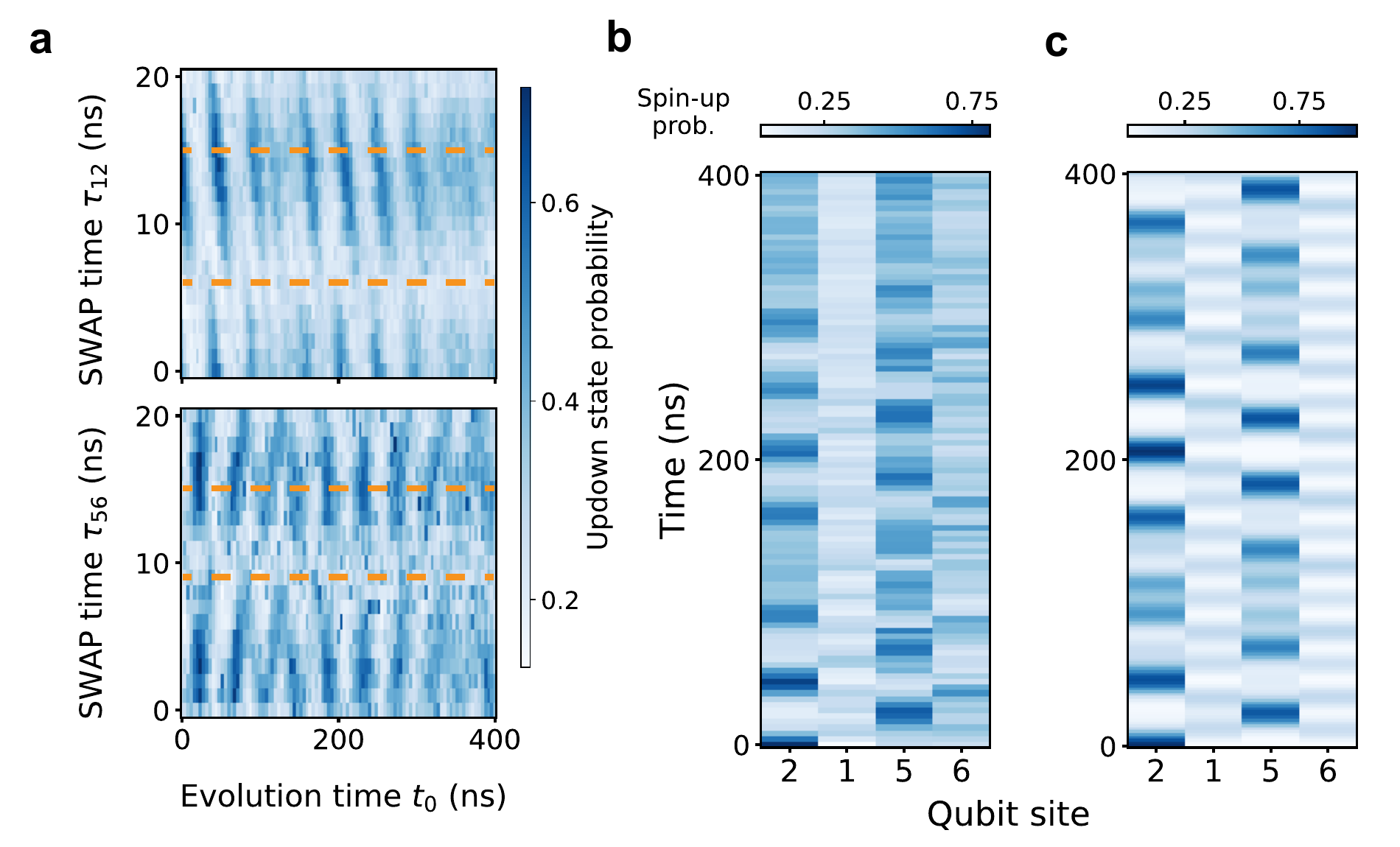}
    \caption{Measurements and simulation of the dynamics of a single excitation $\ket{2} = \ket{\downarrow_1\uparrow_2\downarrow_3\downarrow_4}$ in a 2$\times$2 array with all exchange couplings on and tuned homogeneously, corresponding to Fig. \ref{fig:swaps}i. (a) Unblocked state probability for pair 12 (top panel) and pair 56 (bottom panel) after a variable evolution time $t_0$ and as a function of SWAP time $\tau_{12}$ (top panel) and $\tau_{56}$ (top panel). The readout was performed ramping with an intermediate speed to the PSB point, where the unblocked state is expected to be $\ket{\uparrow\downarrow}$ (see main text). The dashed horizontal lines correspond to rotations by $\pi$ and $2\pi$ (top panel) or $2\pi$ and $3\pi$ (bottom panel), these SWAP times are the same as in Fig. \ref{fig:spin_ring}. These lines correspond to all four single-spin probabilities. (c) Site-resolved quantum walk, where the spin-up probabilities for each site are plotted as a function of evolution time. (d) Simulated quantum walk for an exchange coupling configuration $J_{12} = J_{56} = J_{15} = J_{26} = $\SI{21}{MHz} and using the Zeeman energy values of table \ref{tab:edsr_spectra} (left column), showing good agreement with the measured data. The slight deviation can be explained by a small $g$-tensor modulation that could be caused by the barrier voltage pulses.}
    \label{fig:spin_ring_hom}
\end{figure*}

In Fig. \ref{fig:supp_edsr_spectra}, we report the EDSR lines corresponding to quantum dots 1, 2, 5 and 6 for the magnon experiment. This corresponds to the quantum dots of the left half of the array, at a time where only those four quantum dots were populated. In Table \ref{tab:edsr_spectra}, we list and compare the extracted resonance frequencies with those reported in Fig. \ref{fig:initread}h, when the full array was tuned during a second cooldown. Interestingly, the disorder landscape differs significantly, which can result from different tuning voltages between two different cooldowns, but predominantly from the fact that even within the same cooldown, tuning half of the array or the full array also requires significantly different voltages. As remarked in the main text and observed experimentally (see Fig. \ref{fig:g_factor_mod} below), the $g$-tensor of germanium quantum dots can be electrically modulated.

One important consequence of this change in resonance frequencies is that it crucially changes which spin is initialized as spin-up during the pairwise initialization stage, when the ramps are chosen such that the state $\ket{\uparrow\downarrow}$ is initialized. As outlined in the main text, the spin with the lowest $g$-factor per pair will be initialized in the up-state, while the other will be down. During the first cooldown (left data in Table~\ref{tab:edsr_spectra}), dots 2 and 5 had the lowest $g$-factors, while during the second cooldown (right data), it was dots 1 and 6.

\clearpage

\section{2D plots and quantum walks for static and pair plots}

\begin{figure*}[h]
    \centering
    \includegraphics[width=0.8\textwidth]{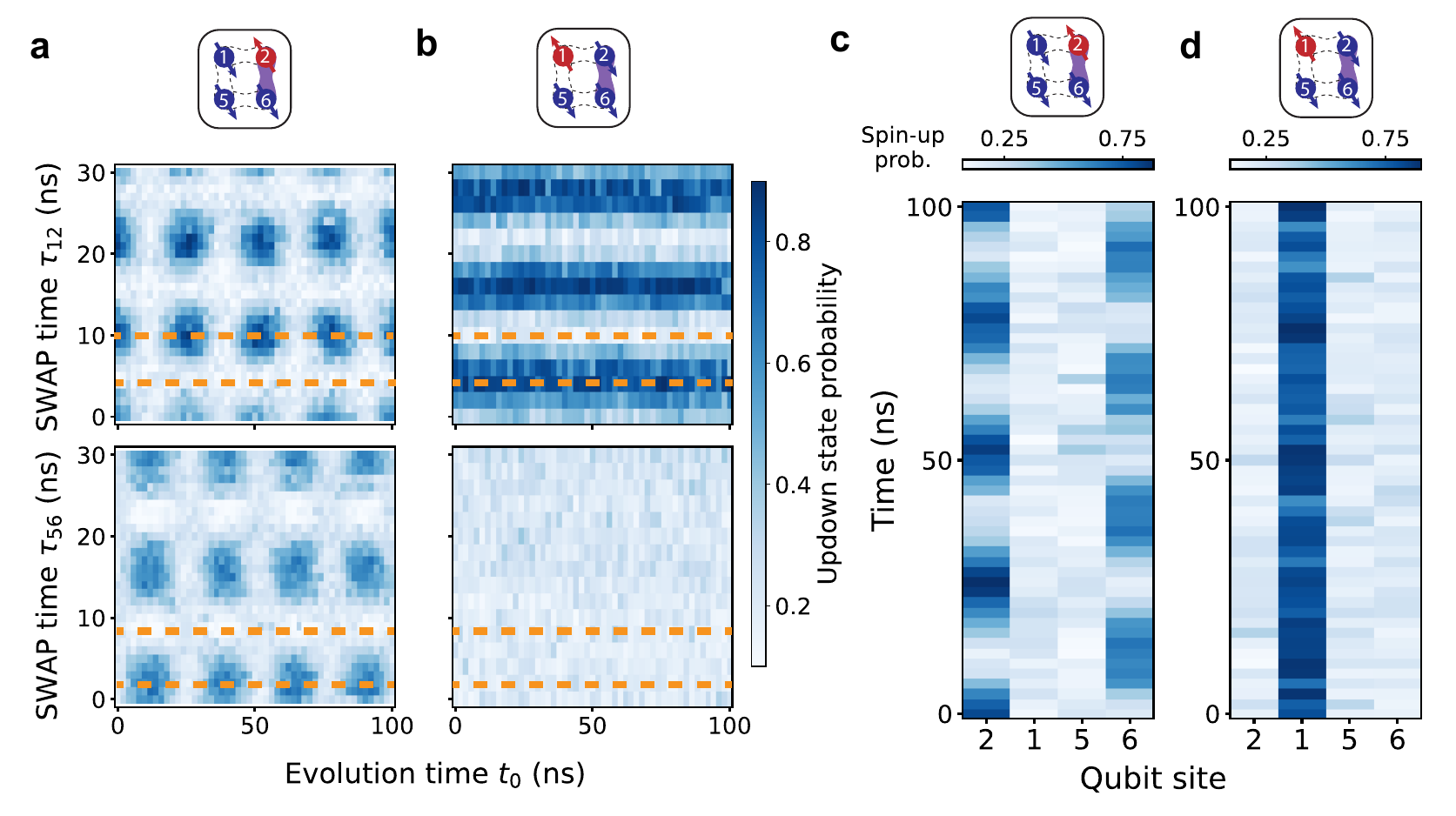}
    \caption{(a) $\ket{\uparrow\downarrow}$ state probability as a function of evolution time and readout SWAP times $\tau_{12}$ and $\tau_{56}$ (see main text), after executing the quantum circuit of panel \ref{fig:swaps}a, but with only $J_{26}$ turned on during the analog evolution phase. The top and the bottom panels correspond to the top (pair 12) and bottom (pair 56) readouts, respectively. (b) Same circuit as for panel (a), but initializing the spin excitation in dot 1. From these 2D maps, we extract the SWAP times which correspond to spin-up readout for each of the four spins (indicated by the dashed orange lines), used to extract the quantum walk data shown in panels (c) and (d). (c) Spin quantum walk corresponding to exchange configuration (a). We observe the magnon oscillate between sites 2 and 6, as expected. (d) Spin quantum walk corresponding to exchange configuration (d), where the magnon remains localized in dot 1. This figure corresponds to Fig. \ref{fig:swaps}g.}
    \label{fig:static_vs_pair}
\end{figure*}

\section{Simulation of quantum walk data}

The quantum walk data simulations reported in the sections below are performed using QuTip. We use the Zeeman energy values reported in Table \ref{tab:edsr_spectra} as well as estimated values of all exchange couplings as input parameters to generate the Hamiltonian matrices of Eq. \ref{eq:spinham} (for the magnon simulations) and Eq. \ref{eq:stham} (for the triplon simulations). Since our spin Hamiltonians are time-independent, we can compute the time evolution by calculating the Hamiltonian's matrix exponential for each time step. Finally, we can compute the single-site magnetization value for each time step and plot the results in the same way as the measured quantum walks. We iterate over various interaction values $\{J_i\}$ around the expected exchange configuration and compare to the measured data.

\clearpage

\section{2D plots, quantum walks and simulation of weakly coupled pairs}

\clearpage

\section{2D plots, quantum walks and simulation of 2$\times$2 ring with inhomogeneous exchange couplings}

\clearpage

\section{2D plots, quantum walks and simulation of 2$\times$2 ring with homogeneous exchange couplings}

\begin{figure*}[h]
    \centering
    \includegraphics[width=0.5\textwidth]{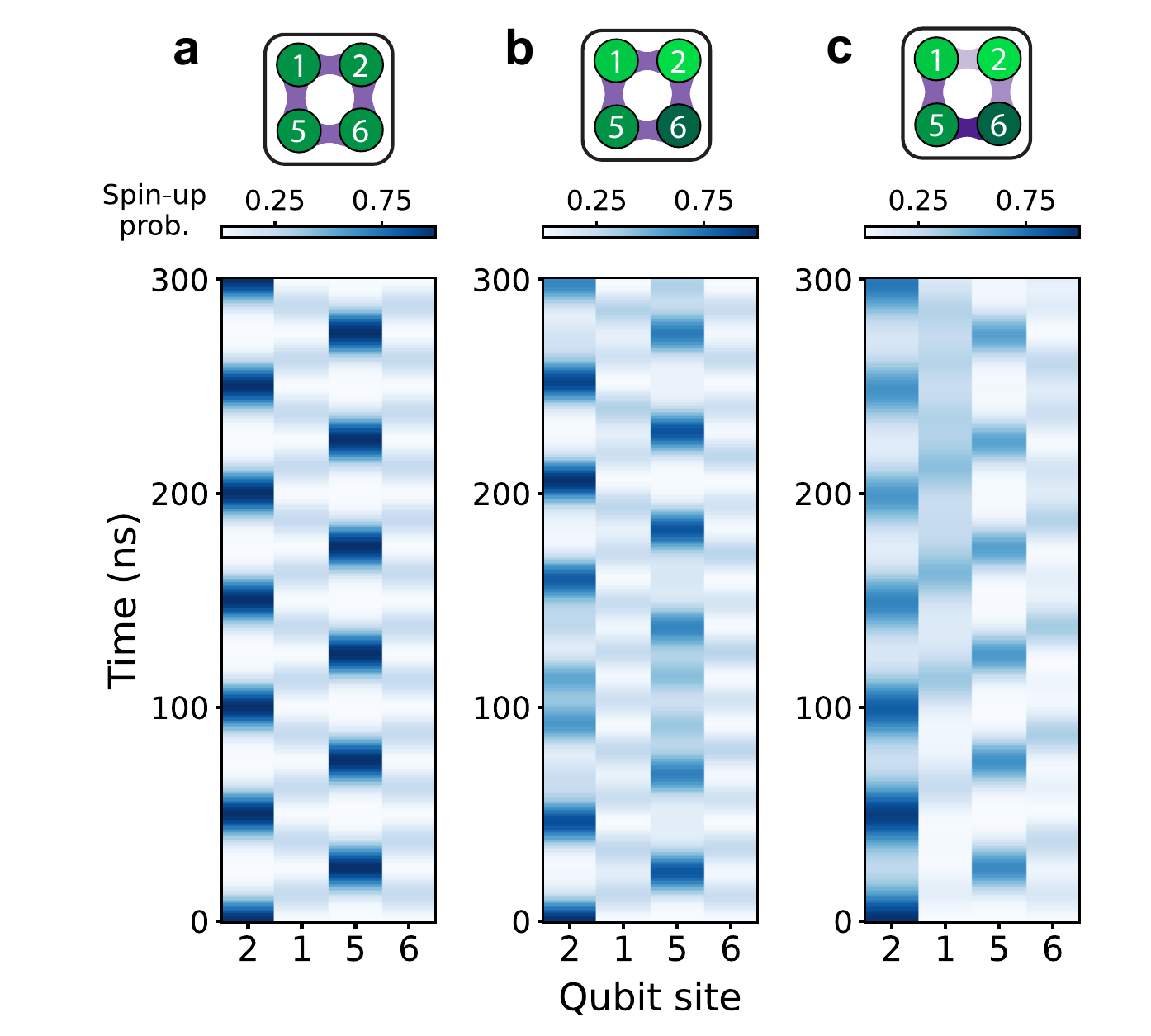}
    \caption{Simulated quantum walk for three different parameter regimes, with a single initial spin excitation in dot 2. (a) All $g$-factors are equal (no single-site disorder) and all exchange couplings are equal $J = $\SI{20}{MHz} (no exchange disorder). The excitation fully refocuses on the other side of the array (dot 5). (b) The Zeeman energies correspond to those measured experimentally, and all exchange couplings are equal $J = $\SI{21}{MHz} (single-site disorder, but no exchange disorder). The obtained pattern is less regular and captures the main features of Fig. \ref{fig:swaps}i, where we aimed to experimentally tune the exchange couplings homogeneously. (c) The Zeeman energies correspond to those measured experimentally, and the input exchange interactions are $J_{12} = $\SI{11.5}{MHz}, $J_{56} = $\SI{27.0}{MHz}, $J_{15} = $\SI{20.0}{MHz} and $J_{26} = $\SI{15.5}{MHz} (single-site disorder and exchange disorder). The obtained pattern captures the main features of Fig. \ref{fig:swaps}j.}
    \label{fig:spin_ring_sim}
\end{figure*}

\begin{figure*}[h]
    \centering
    \includegraphics[width=0.9\textwidth]{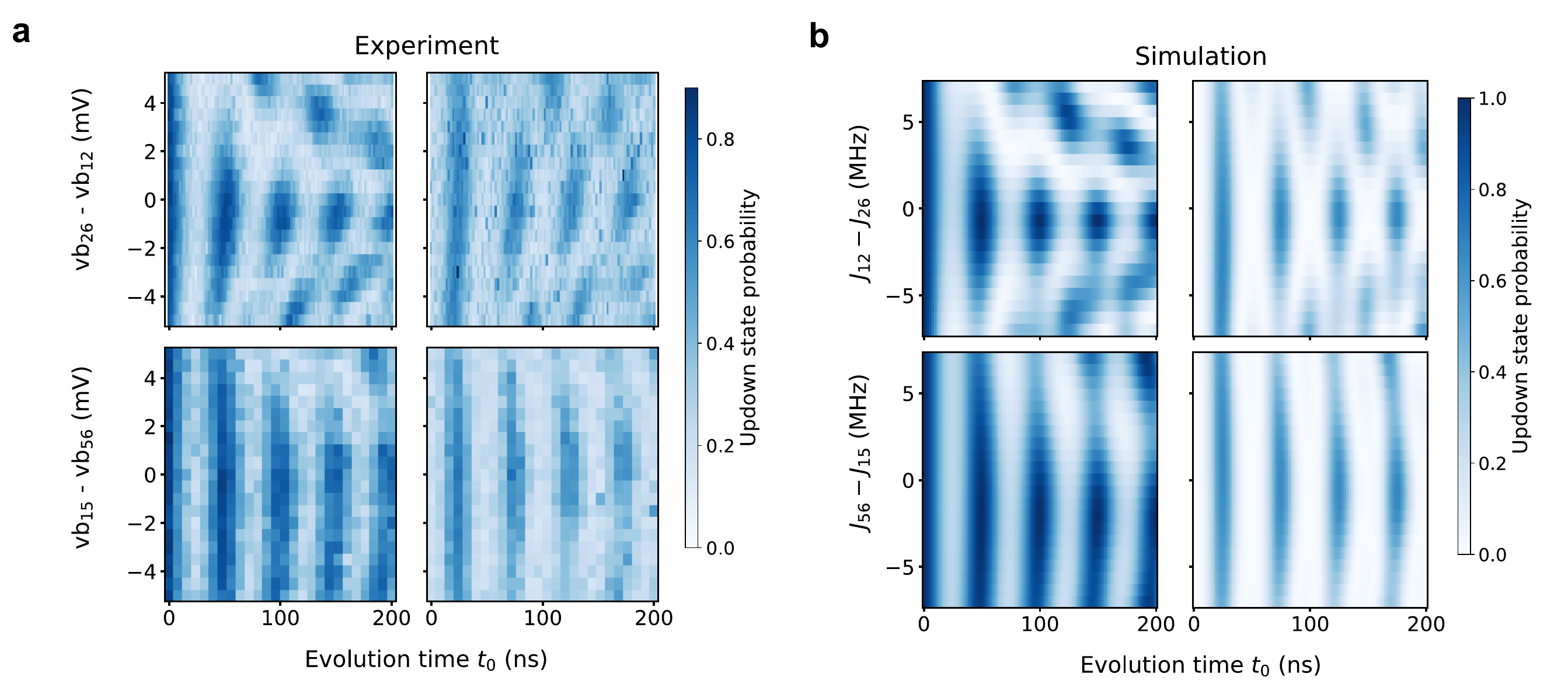}
    \caption{Dynamics of a single excitation $\ket{2} = \ket{\downarrow_1\uparrow_2\downarrow_3\downarrow_4}$ in a 2$\times$2 array with all exchange couplings on, corresponding to the quantum walk of Fig. \ref{fig:swaps}j, as a function of barrier voltage combinations vb$_{26}$ - vb$_{12}$ (top two panels) and vb$_{15}$ - vb$_{56}$ (bottom two panels). For each row, the first plot corresponds to the readout of pair 12, the second plot to the readout of pair 56. The relative voltage value of 0 corresponds to the voltage configuration for which Fig. \ref{fig:swaps}j was taken. Away from that center point, the oscillations (corresponding to the spin-up probabilities of dots 2 and 5, respectively) become less regular and show a more complex pattern. (b) Simulation of the data in (a), starting at the exchange configuration of Fig. \ref{fig:spin_ring_sim}c, showing good agreement with the experimentally observed oscillation patters. This supports the correctness of the previously simulated exchange interaction values.}
    \label{fig:spin_ring_2D}
\end{figure*}

\clearpage

\section{8-spin quantum walk of four uncoupled pairs}

\begin{figure*}[h]
    \centering
    \includegraphics[width=0.75\textwidth]{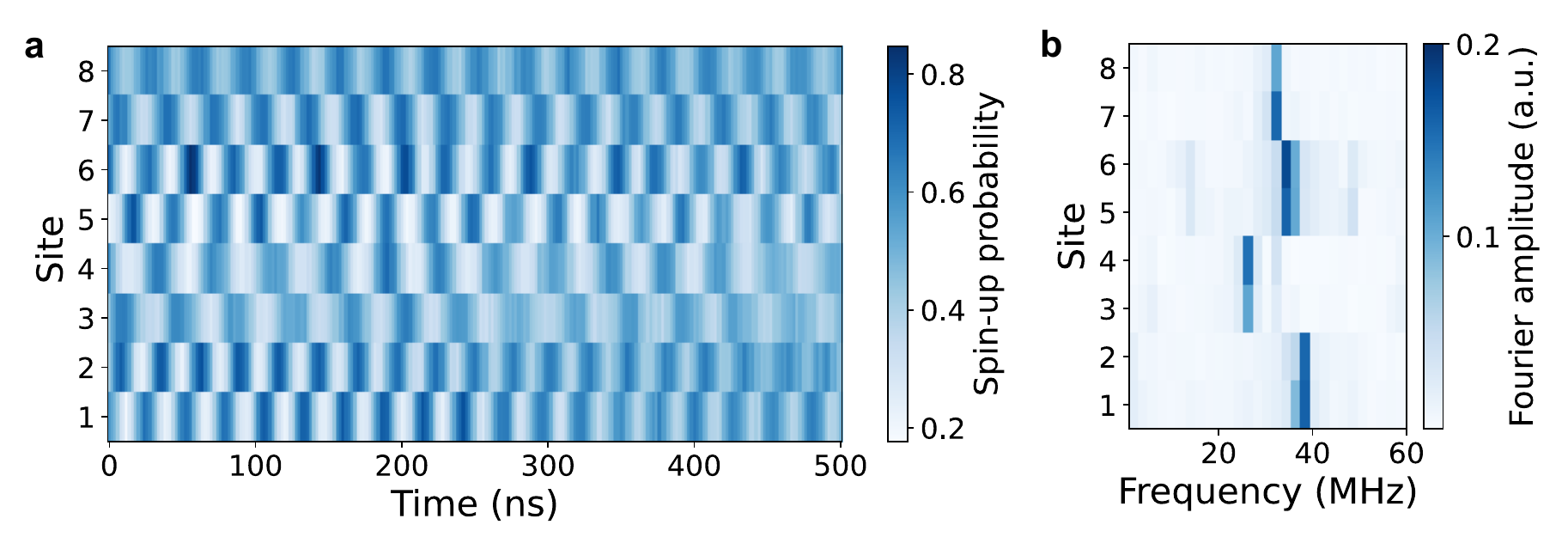}
    \caption{(a) 8-site quantum walk with staggered couplings, resulting in four uncoupled pairs. The initial state is a Néel state (up to a single initial SWAP for pair 34). The evolution pattern is consistent with four uncoupled pairs, where each spin excitation does not leave the pair it was initialized in. This plot demonstrates the ability to simultaneously initialize all spins in the array. State readout was performed in two separate experimental rounds. In the first round, all spin pairs were read out sequentially, resulting in four bits of information, corresponding to the spin-up probability of the spin in each pair with the lowest $g$-factors (see main text). In the second round, a SWAP gate is applied on each pair before readout, allowing us to obtain the spin-up probability of the other dot. Note that, for some pairs, a small phase mismatch is observed for longer evolution times, which we attribute to imprecise SWAP gate calibration.  (b) Fourier transform of (a). For each quantum dot pair, the oscillation frequency is the same for the two spins, consistent with the spins only oscillating between the dots of the same pair.}
    \label{fig:8spins_Neel}
\end{figure*}

\clearpage

\section{Measured initialization data corresponding to Fig. \ref{fig:swaps}l}

\begin{figure*}[h]
    \centering
    \includegraphics[width=0.95\textwidth]{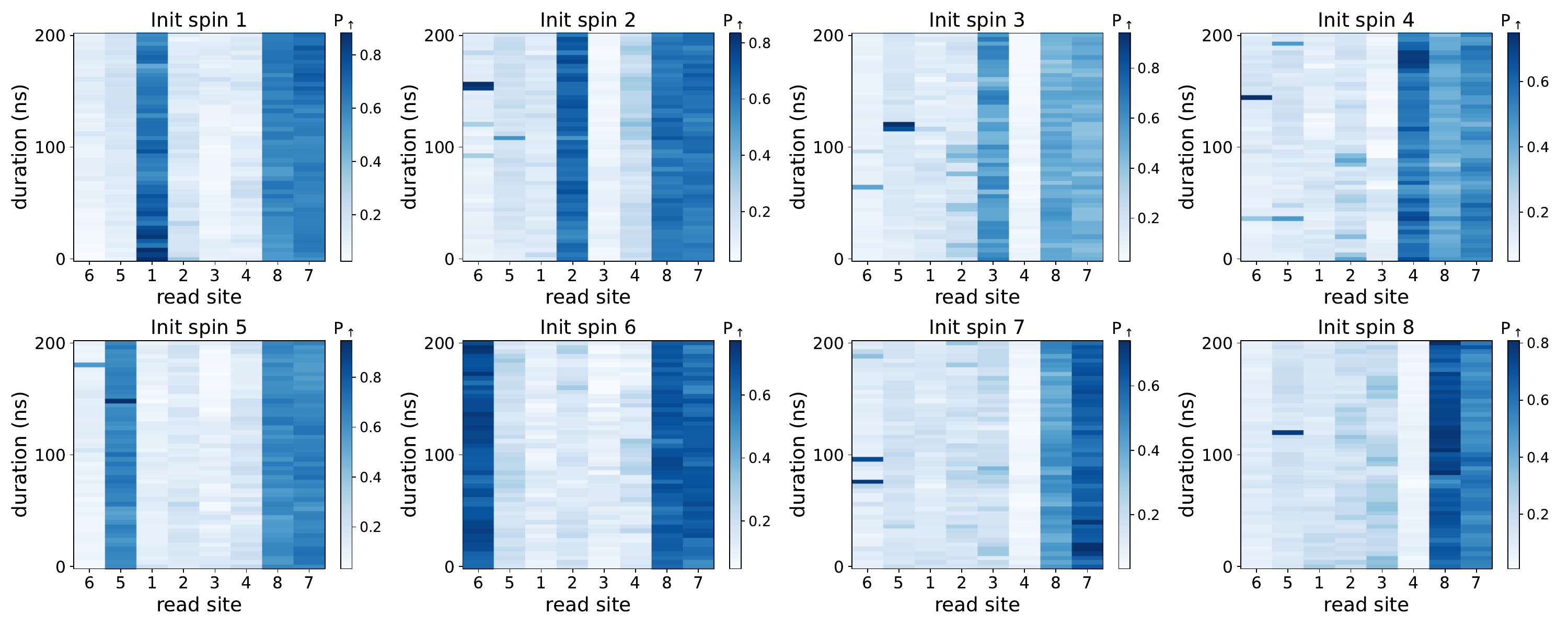}
    \caption{Static quantum walk plots for single magnons initialized in different locations in the array, corresponding to the data shown in Fig. \ref{fig:swaps}l after averaging over the y-axis. For each plot, the state $\ket{2} = \ket{\downarrow_1\uparrow_2\downarrow_3\downarrow_4\downarrow_5\downarrow_6\downarrow_7\downarrow_8}$ was initialized, with a single magnon in dot 2. Subsequently, a sequence of SWAP gates was applied to move the excitation around, as depicted on the quantum circuit of Fig. \ref{fig:swaps}k. For dots 1-6, the measurement results in the expected initialized spin-up. For dot pair 7-8, the readout visibility was very low at the time of measurement, resulting in a seemingly always high signal.}
    \label{fig:sup_init_8spins}
\end{figure*}

\clearpage

\section{EDSR, exchange splitting and $g$-factor modulation}

\begin{figure*}[h]
    \centering
    \includegraphics[width=0.75\textwidth]{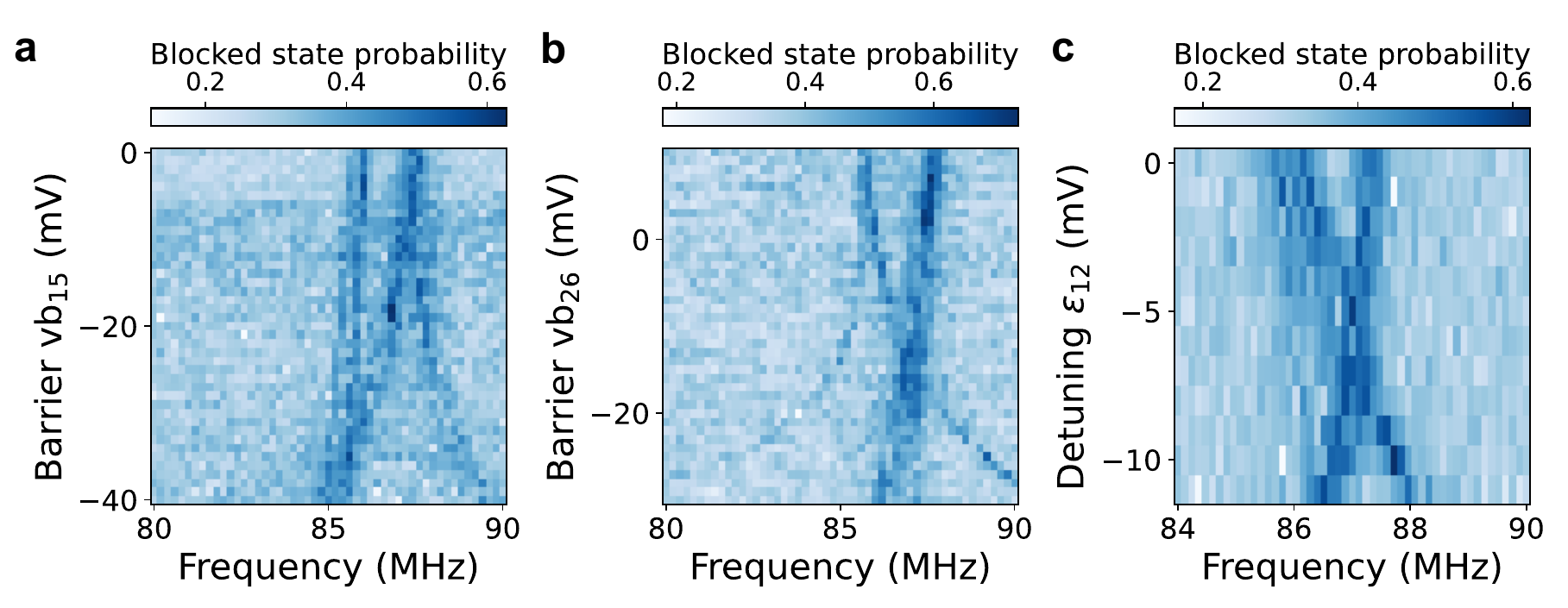}
    \caption{Microwave spectroscopy of Zeeman energies, exchange control and $g$-factor modulation. (a) Resonances as a function of barrier vb$_{15}$. Exchange splitting is only observed for the rightmost resonance, which corresponds to dot 1. The splitting is a direct measure of $J_{15}$. (b) Resonances as a function of barrier vb$_{26}$. Exchange splitting is only observed for the leftmost resonance, which corresponds to dot 2. The splitting is a direct measure of $J_{26}$. An additional splitting is observed for dot 1, which we attribute to bad virtualization of $J_{15}$ with respect to barrier voltage vb$_{26}$. (c) EDSR resonance of spins 1 and 2 as a function of detuning $\varepsilon_{12}$. A $g$-factor crossing is observed at $\varepsilon_{12} \approx$ \SI{6}{mV}. }
    \label{fig:g_factor_mod}
\end{figure*}

In this work, we have used EDSR as a method to spectroscopically determine the resonance frequency of each spin. Since we obtain two different resonances for each quantum dot pair, we need to identify which resonance corresponds to which spin within each pair. Fig. \ref{fig:g_factor_mod}a and b show the resonance frequencies of Q1 and Q2 at a magnetic field of \SI{20}{mT}, and as a function of barrier voltages vb$_{15}$ and vb$_{26}$, respectively. Spins 5 and 6 are not initialized for this experiment and thus remain in a mixed state. We observe the expected branching of the resonance frequencies as a function of exchange interaction. Given that the rightmost resonance branches with increased $J_{15}$ (Fig. \ref{fig:g_factor_mod}a), we infer that this frequency corresponds to dot 1. Similarly, $J_{26}$ causes the branching of the leftmost resonance (Fig. \ref{fig:g_factor_mod}b), which we then attribute to dot 2. Thus, this straightforward technique allows us to identify which resonance corresponds to which spin for each quantum dot pair, which is relevant especially for the initialization of the state $\ket{\uparrow\downarrow}$, where the spin-up is initialized (and read out) on the spin with the lowest Zeeman energy. This data corresponds to the first cooldown, where only the left half of the array was tuned (see Fig. \ref{fig:supp_edsr_spectra}).

Additionally, we can demonstrate $g$-factor tunability for this quantum dot pair exemplarily, where we scan the detuning $\varepsilon_{12}$ as we record the position of both resonance frequencies (Fig \ref{fig:g_factor_mod}c). Strikingly, for this quantum dot pair, the Zeeman energies are modulated such that they cross at negative detunings. This is not observed for each pair: the tunability of the $g$-factors relies on microscopic details such as local disorder in the heterostructure and the shape of the quantum dot wavefunction.

\clearpage

\section{Theory of exchange-coupled singlet-triplet qubits}

We consider the Hamiltonian of Eq. \ref{eq:spinham} describing an array of exchange-coupled spins. As outlined in the main text, for this discussion it suffices to consider this isotropic Hamiltonian and neglect spin-orbit terms (a full derivation of the two-qubit Hamiltonian of two singlet-triplet qubits can be found in \cite{zhang_universal_2024}). Projecting this Hamiltonian onto the S-T$^{-}$ basis of two qubits, spanned by $\{\ket{SS}, \ket{ST^{-}}, \ket{T^{-}S}, \ket{T^{-}T^{-}}\}$, we obtain the Hamiltonian of Eq. \ref{eq:stham} \cite{zhang_universal_2024} for two qubits. Rewritten so that we treat interaction and disorder terms separately, this Hamiltonian reads:

\begin{equation} \label{eq:stham_sup}
\begin{split}
    H_{ST,\text{2Q}} &= \frac{1}{2}\left(\bar{E}_{z,1}-J_1^\perp-\frac{J_{12}^\parallel}{4}\right) \sigma_1^z \\ 
    &+ \frac{1}{2}\left(\bar{E}_{z,2}-J_2^\perp-\frac{J_{12}^\parallel}{4}\right) \sigma_2^z \\ 
    &+\frac{J_{12}^\parallel}{4} \left( \sigma_i^x \sigma_j^x + \sigma_i^y \sigma_j^y
    + \frac{1}{2} \sigma_i^z\sigma_j^z\right),
\end{split}
\end{equation}

where the first two terms represent the single-site disorder, and the last term is the qubit-qubit interaction. For a two-site system, tuning to the homogeneous condition is equivalent to finding $J_1^\perp$ and $J_2^\perp$ such that $\bar{E}_{z,1}-J_1^\perp = \bar{E}_{z,2}-J_2^\perp$. However, when considering a one-dimensional chain of $N$ sites, also the individual, parallel exchange interactions need to be considered. Even if all parallel interactions are tuned equally, i.e. $J_{12}^\parallel = J_{23}^\parallel = ... = J_{N-1,N}^\parallel = J^\parallel$, the first and last site would have a different disorder contribution due to the absence of periodic boundary conditions. For $N=3$ qubits, the corresponding homogeneous condition reads:

\begin{equation} \label{eq:stham_cond}
\begin{split}
     \bar{E}_{z,1}-J_1^\perp-\frac{J_{12}^\parallel}{4} = \bar{E}_{z,2}-J_2^\perp-\frac{J_{12}^\parallel}{4}-\frac{J_{23}^\parallel}{4} = \bar{E}_{z,3}-J_3^\perp-\frac{J_{23}^\parallel}{4}
\end{split}
\end{equation}

The calibration procedure outlined in the main text allows for the independent and sequential calibration of all $J_i^\perp$ terms, allowing us to achieve the condition above for any values of $J_{12}^\parallel$ and $J_{23}^\parallel$. For the regime where both parallel exchanges are equal, which we tuned up for the measurements of Fig. \ref{fig:qwalks_st}d-e, the final interaction Hamiltonian is an XXZ Hamiltonian with an anisotropy parameter $\Delta = 0.5$, which simply reads:

\begin{equation} \label{eq:stham_XXZ}
H_{\text{3Q,int}} = \frac{J^\parallel}{4} \sum_{i=1}^2 \left( \sigma_i^x \sigma_{i+1}^x + \sigma_i^y \sigma_{i+1}^y + \frac{1}{2} \sigma_i^z\sigma_{i+1}^z \right).
\end{equation}

Given the iterative nature of the calibration procedure, this method can be scaled in principle to obtain the Hamiltonian above for any $N$-site chain of singlet-triplet qubits.

\begin{figure*}[h]
    \centering
    \includegraphics[width=0.43\textwidth]{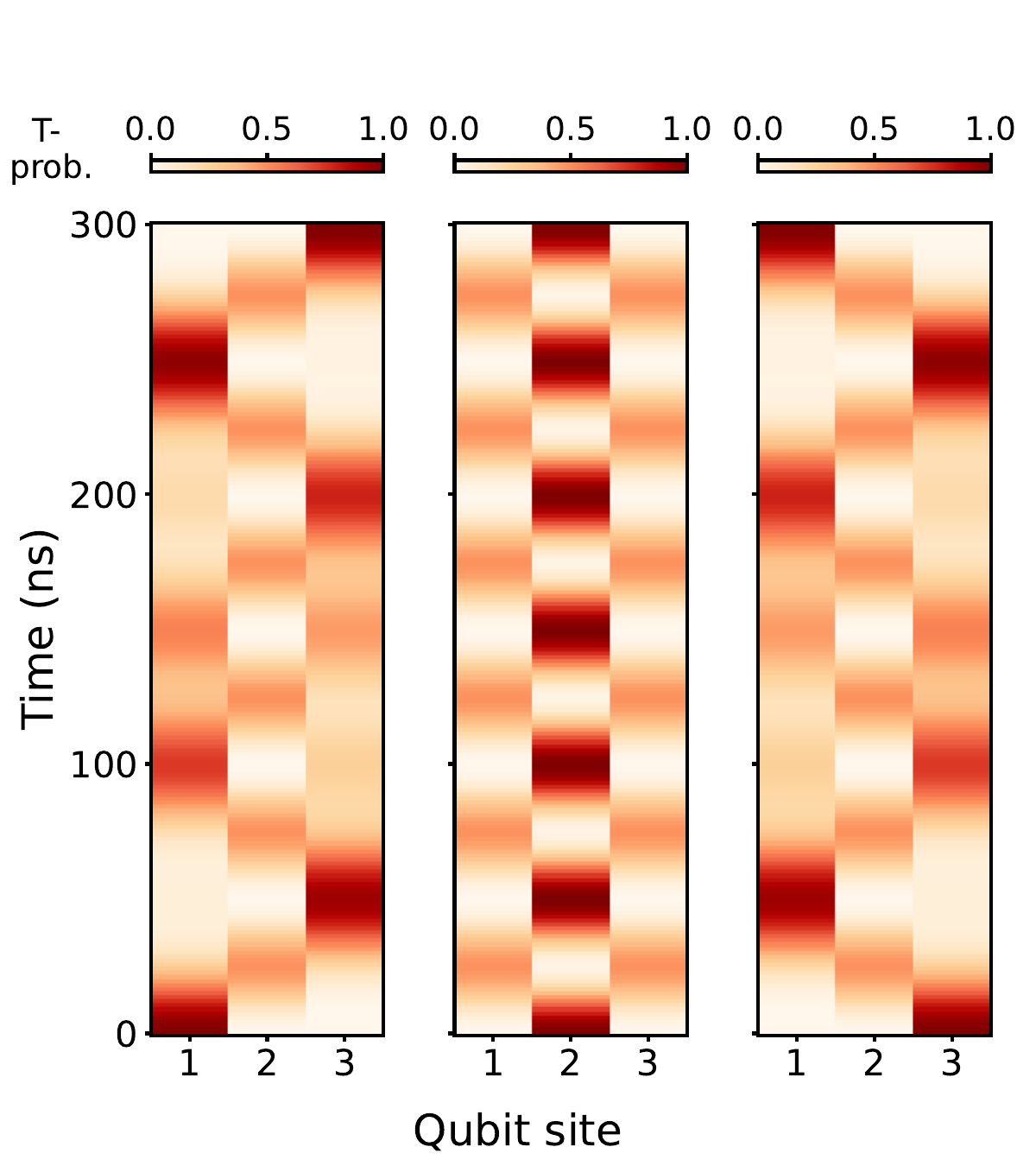}
    \caption{Simulated three-site quantum walk plot of a triplon excitation initialized in Q1, Q2 and Q3, respectively. For this simulation, we let each initial state evolve under the Hamiltonian of Eq. \ref{eq:stham_XXZ} and we assume equal parallel exchange couplings $J^\parallel = $ \SI{14}{MHz}. This simulation is in excellent agreement to the measured triplon quantum walks of Fig. \ref{fig:qwalks_st}e.}
    \label{fig:st_qwalk_sim}
\end{figure*}

We use Eq. \ref{eq:stham_XXZ} to simulate the triplon propagation of Fig. \ref{fig:qwalks_st}e, which is plotted in Fig. \ref{fig:st_qwalk_sim}, and find excellent agreement with the measured propagation pattern. The simulation yields an exchange interaction $J^\parallel = $ \SI{14}{MHz}. The difference in visibility of the experimental data and the simulations can be explained by the difference in readout fidelity for Q1, Q2 and Q3. Also, small differences can be attributed to slightly different values of $J_{12}^\parallel$ and $J_{23}^\parallel$.

\clearpage

\section{Correlation plots for triplon quantum walks}

\begin{figure*}[h]
    \centering
    \includegraphics[width=0.65\textwidth]{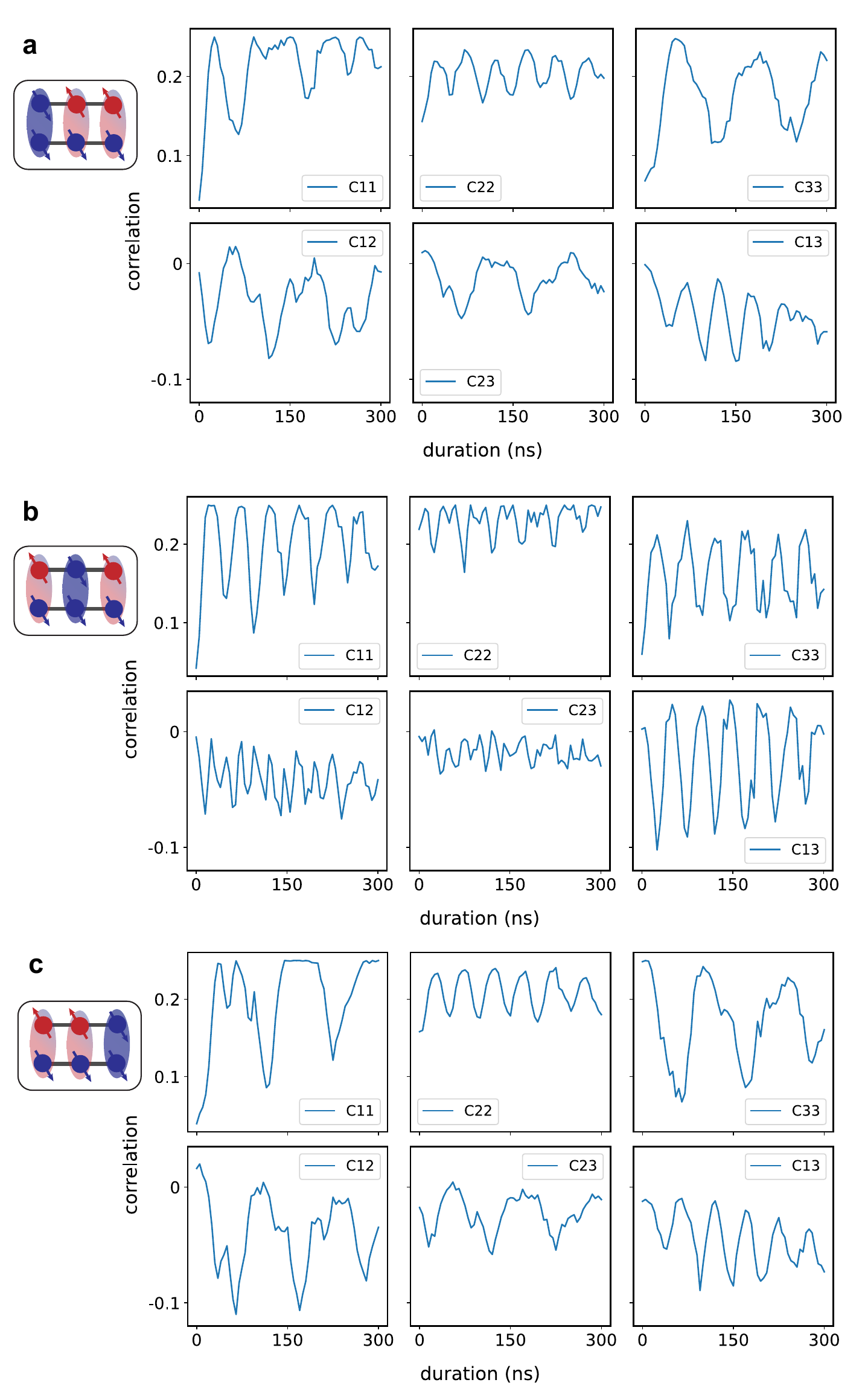}
    \caption{Site-site correlations $C_{ij}$ for every pair of sites $(i,j)$, corresponding to the quantum walks of \ref{fig:qwalks_st}e, where a triplon excitation was initialized in site 1, 2 or 3, respectively.}
    \label{fig:ST_corr}
\end{figure*}

Fig. \ref{fig:ST_corr} shows all site-site correlations $C_{ij}$ corresponding to the quantum walk data of Fig. \ref{fig:qwalks_st}e. When $i \neq j$, we observe the correlation plots become negative for all combinations of $i$ and $j$ and irrespective of the site in which the triplon is initialized. This is consistent with the propagation of the triplon across the array and the generation of entanglement. Note that the magnitudes of the correlation values remain still relatively small, mainly due to the reduced readout fidelity. 

\clearpage

\section{Simulation of ST$^-$ two-site phase diagram}

\begin{figure*}[h]
    \centering
    \includegraphics[width=0.4\textwidth]{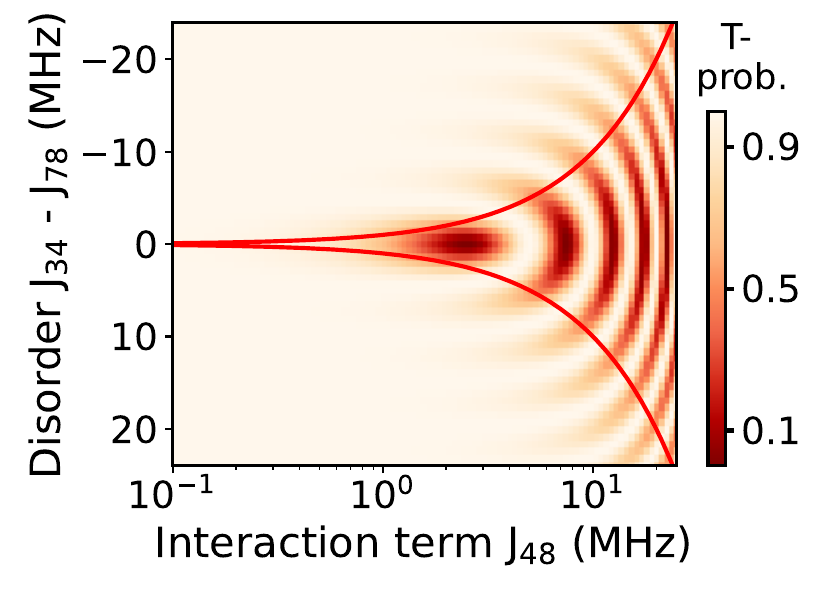}
    \caption{Simulated evolution of an initial state $\ket{ST^{-}}$ under the Hamiltonian of Eq. \ref{eq:stham}, and corresponding to the measurement of Fig. \ref{fig:qwalks_st}g. The x-axis corresponds to the coupling term $J_{\parallel} = J_{48}$, while on the y-axis, the difference in single-site disorder is scanned using the exchange couplings $J_{\perp,1} = J_{34}$ and $J_{\perp,2} = J_{78}$. The simulated pattern is in good agreement with the experimental results. In addition, the red line corresponds to the points where the single-site disorder is equal to the coupling strength.}
    \label{fig:st_phase_diag}
\end{figure*}

In Fig. \ref{fig:st_phase_diag}, we simulate the evolution of an initial state $\ket{ST^{-}}$ of a two-site system subject to the Hamiltonian of Eq. \ref{eq:stham}, where $J_{\perp,1} = J_{34}$, $J_{\perp,2} = J_{78}$ and $J_{\parallel} = J_{48}$, corresponding to the labels used in section \ref{sec:S-T$^{-}$_triplons} and Fig. \ref{fig:qwalks_st}g. As in the experiment, the initial state is allowed to evolve for \SI{200}{ns}, and we vary both the interaction as well as the disorder terms independently. In the simulation, for each combination of exchange terms, we retrieve the magnetization for the first site, where the exchange interactions are scanned in an exponential way which corresponds to a linear scan in barrier gate voltage. The resulting pattern agrees very well with Fig. \ref{fig:qwalks_st}g and is consistent with a variation of coupling strength of more than two orders of magnitude. The red line corresponds to the points where the coupling and the disorder terms are equal, serving as a ``phase boundary" between the regions of high interaction-to-disorder ratio, where the excitation can freely evolve, and that of low interaction-to-disorder ratio, where the excitation remains localized. Note that, for simplicity, we set the average Zeeman energy of both singlet-triplet qubits to \SI{50}{MHz}, which is approximately the real experimental values. 

\clearpage

\section{Triplon quantum walk for four sites}

\begin{figure*}[h]
    \centering
    \includegraphics[width=0.9\textwidth]{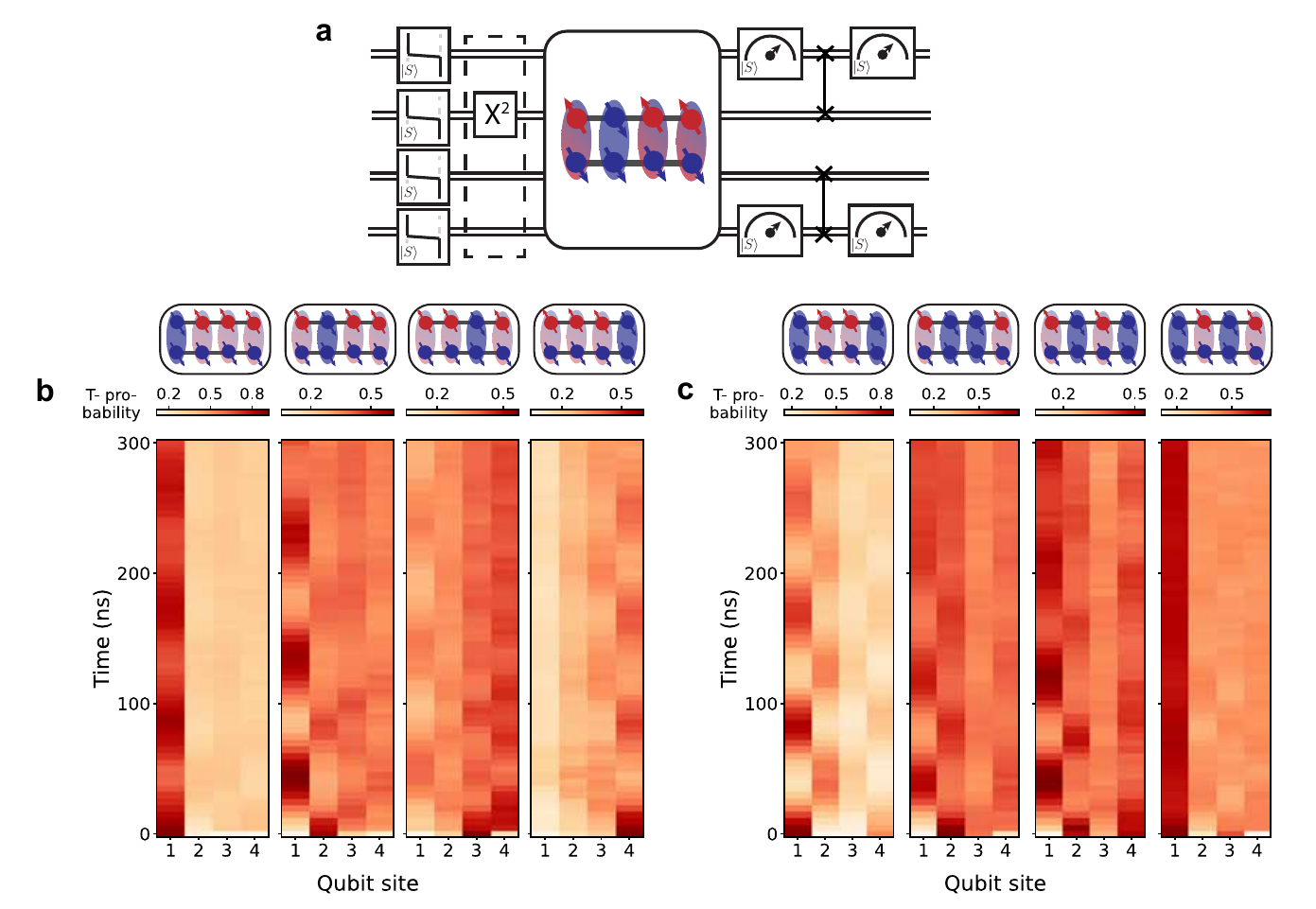}
    \caption{Triplon quantum walk for four coupled singlet-triplet qubits. (a) Circuit diagram showing the state preparation, evolution and readout steps for the 4-site triplon quantum walk. Readout is performed in two steps, with a SWAP gate in between to transfer the population of Q2 (Q3) to Q1 (Q4). (b) Quantum walk results for a single excitation initialized in sites 1, 2, 3 or 4, respectively. Note that at the point of the experiment, the readout fidelity of the right side of the device was much lower than for the left side, resulting in a lower visibility for the readout of Q3 and Q4. We do not correct for the difference in readout visibility. (c) Quantum walk results for two excitations in the array and for different initial states.}
    \label{fig:ST_qwalk_4sites}
\end{figure*}

Fig. \ref{fig:ST_qwalk_4sites} shows the measured triplon quantum walk data for a chain of four coupled singlet-triplet qubits, i.e. using all eight spins of the 2$\times$4 ladder, with the circuit diagram corresponding to the measurements depicted in Fig. \ref{fig:ST_qwalk_4sites}a. The data shown corresponds to the initialization of either a single triplon (\ref{fig:ST_qwalk_4sites}b) or two triplons (\ref{fig:ST_qwalk_4sites}c). In this case, the visibility of the oscillations seems to vary strongly between the different panels, suggesting either inhomogeneities in the exchange coupling configuration or, most likely, strong differences in readout visibility for each site.

\end{document}